\documentclass[journal]{IEEEtran}

\usepackage[english]{babel}

\usepackage{amsmath}
\usepackage{subcaption,graphicx}
\usepackage[colorlinks=true, allcolors=blue]{hyperref}
\usepackage{caption}
\usepackage{subcaption}
\usepackage[noadjust]{cite}

\title{Wide Band Interaction Impedance and Mode Excitation in Glide Symmetric Double Corrugated Waveguides for mm-wave TWTs}

\author{\IEEEauthorblockN{
Nelson Castro\IEEEauthorrefmark{1},   
Miguel Saavedra-Melo\IEEEauthorrefmark{2},   
Eva Rajo-Iglesias \IEEEauthorrefmark{1},    
Filippo Capolino\IEEEauthorrefmark{2}\IEEEauthorrefmark{1}    
}                                     
\\
\IEEEauthorblockA{\IEEEauthorrefmark{1}
Department of Signal
Theory and Communications, University Carlos III of Madrid, Leganés, Spain}
\IEEEauthorblockA{\IEEEauthorrefmark{2}
 Department of Electrical Engineering and Computer Science, University of California, Irvine, California, USA}
}
\begin{document}


\maketitle

\begin{abstract}
Focusing on traveling wave tube (TWT) applications, the interaction impedance between an electron beam and electromagnetic modes in three distinct, but related, corrugated waveguides that operate at millimeter waves is investigated together with the role of glide symmetry. Two waveguide structures have glide symmetry, and the irreducible Brillouin zone is related to half of the period, leading to a wide band linearity, i.e., nondispersive, property of the dispersion diagram. The investigation on the modes with longitudinal electric field that can be excited shows that the bottom-top glide (BT Glide) symmetric corrugated waveguide has a wide band interaction impedance, hence it is a good candidate for millimeter wave TWT amplifiers. Furthermore, the backward electromagnetic mode in such BT Glide slow wave structure is not $z$ polarized,  eliminating the {risk} of backward wave oscillations. 
\end{abstract}
\section{Introduction}

Slow Wave Structures (SWS) are RF structures where the electromagnetic wave is confined and the phase velocity is less than the speed of light. In a traveling wave tube (TWT), an input signal is amplified due to the synchronous interaction between the slow wave within a periodic structure and an electron beam along a narrow vacuum channel \cite{pierce-twt50, schachter11CH1}.
Research on TWT amplifiers has increased considering their ability to generate power at millimeter waves  \cite{paoloni21,Booske2008,Sengele2009,booke11, Armstrong2012,joye14,Armstrong2018,Armstrong2020}. 
Some TWTs have very wide instantaneous bandwidth, which allows signals to be amplified linearly up to the kilowatt level, even in the millimeter wavelength (mmW) bands \cite{joye14}, where TWTs are excellent devices for {high-data-rate} communication systems, especially when broadband transmission with moderate power amplification {is required, as highlighted in} \cite{chong13}. 
The wide bandwidth capability makes them ideal for applications such as communications, advanced radar, military electronic countermeasure systems, and space research \cite{booke11}. 


{In the millimeter wave and sub-THz bands, the most popular SWS is the serpentine waveguide which can provide wide bandwidth, good coupling, and relatively high interaction impedance \cite{paoloni21}. A detailed discussions of its properties and design methodology is in \cite{Nguyen2014,joye14}. In recent years, straight rectangular SWSs have been introduced which are relatively easy to fabricate since the beam tunnel may be easier to integrate with the structure, as it can be seen in \cite{baig17,mineo10,paoloni13,paoloni14,Basu22}.}

\begin{figure*}[htpb]
     \vspace{-0.6cm}
     \centering
     \begin{subfigure}[htpb]{0.3\textwidth}
         \centering
    \includegraphics[width=\textwidth]{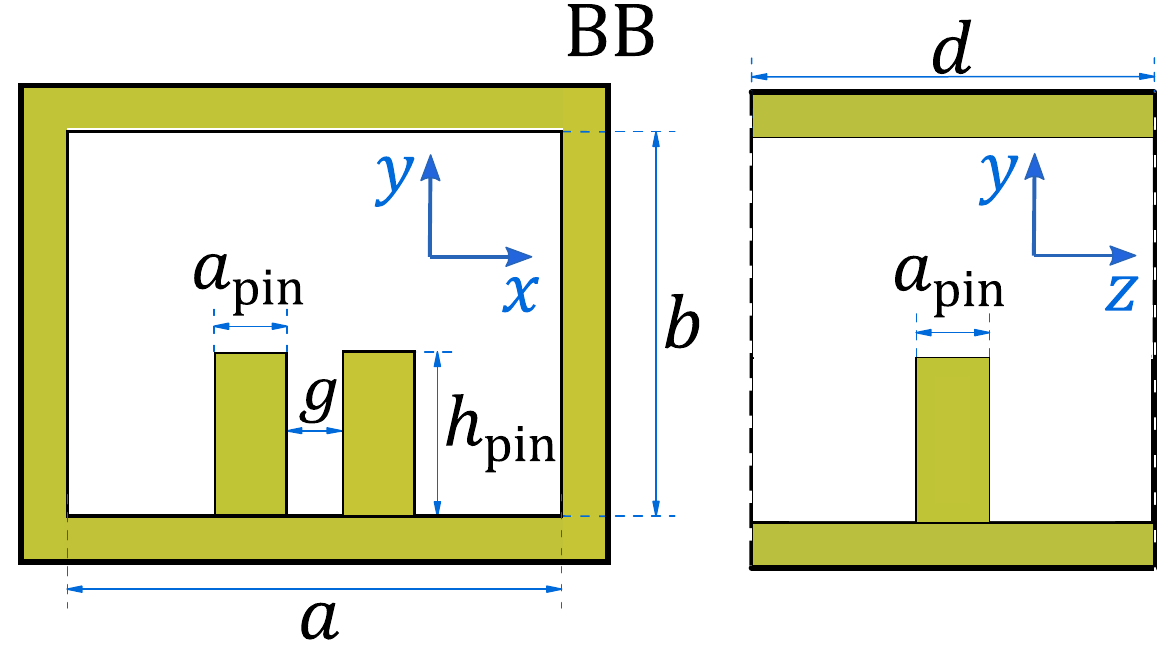}
         \caption{}
         \label{fig:BB_uc}
     \end{subfigure}
     \hfill
     \begin{subfigure}[htpb]{0.3\textwidth}
         \centering
         \includegraphics[width=\textwidth]{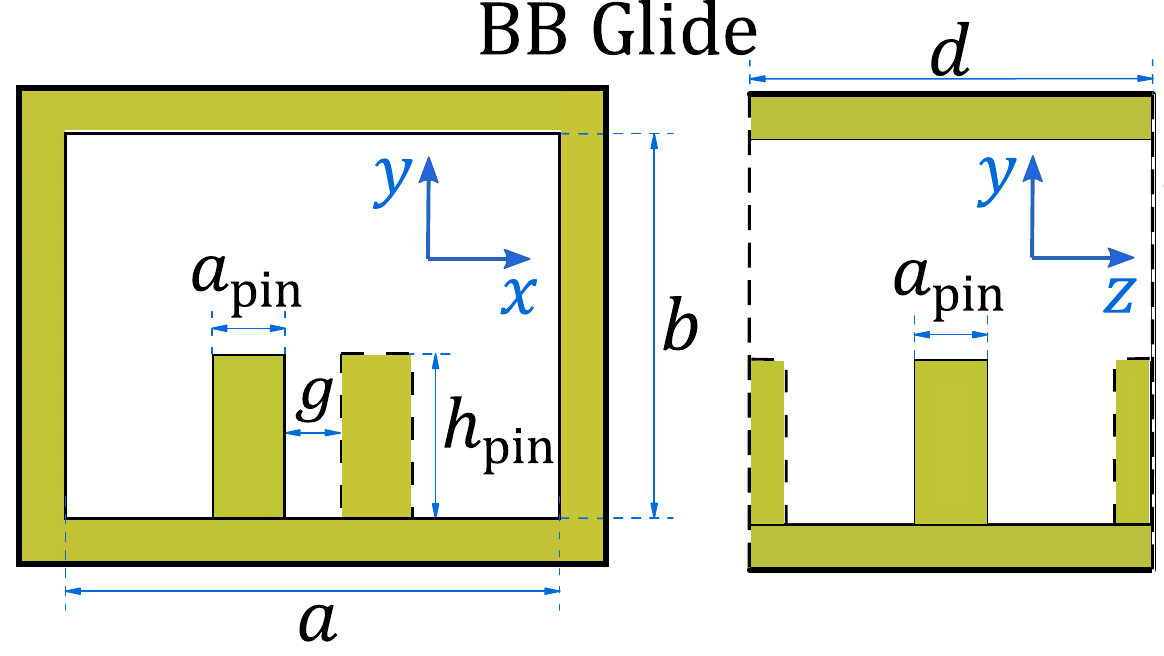}
         \caption{}
         \label{fig:BB_glide_uc}
     \end{subfigure}
     \hfill
     \begin{subfigure}[htpb]{0.3\textwidth}
         \centering
         \includegraphics[width=\textwidth]{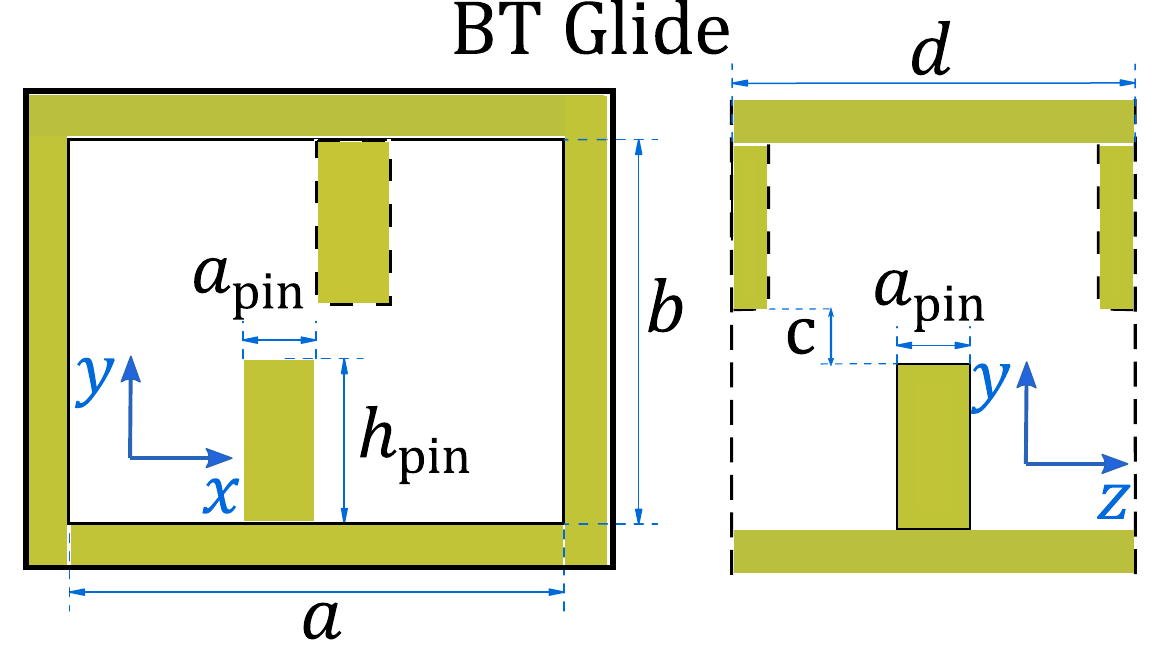}
         \caption{}
         \label{fig:BT_glide_uc}
     \end{subfigure}
        \caption{Unit cells of the three SWSs supporting modes propagating in the $z$ direction, made of a metallic waveguide with periodic pillars (the corrugation of period $d$) of square cross-section: (a) Bottom-bottom corrugations (BB);  (b) Bottom-bottom corrugations with glide symmetry (BB Glide); (c) Bottom-top corrugations with glide symmetry (BT Glide).}
        \label{fig:unit-cells}
\end{figure*}

Glide symmetry in periodic structures like the TWT was originally studied in \cite{crepeau64,Mittra65,Hessel73} and has gained notable interest as this feature has several advantages, such as low dispersion, wide stopbands, and high levels of on-axis anisotropy (used in lens antenna design) \cite{quevedo21}. Glide symmetry was used in electron-beam devices for long time including the 1950s \cite{Gould58, Harvey60,Kieburtz70,Staprans73} but without knowing the deep spectral properties later highlighted in \cite{Hessel73}. 
This type of high symmetry, as well as screw symmetry, has been used in microwave structures such as filters, traveling wave tubes, and traveling-wave antennas \cite{Hessel73}. Recently, interest in glide symmetry has been resurrected with new and fascinating applications \cite{quevedo21,quevedo20}, such as widening the bandwidth and increasing the attenuation of holey structures at the stopband, useful for producing cost-effective gap waveguides \cite{ebrahimpouri18}. Regarding recent TWTs, straight rectangular SWSs with glide symmetric staggered vane interaction structure have been previously studied (see \cite{Field18} and references therein), leading to, for example, a 100-W, 200-GHz high bandwidth amplifier. Furthermore, an offset double corrugated SWS as enabling power amplifier for long range sub-THz links was proposed in \cite{Basu2021} and in \cite{Patent} where the pillars (i.e, corrugations) exhibit glide symmetry (a structure here referred to as "BB Glide", see Fig. \ref{fig:unit-cells}), though no comprehensive data and results were shown. 

\begin{figure*}
     \centering
     \begin{subfigure}[htpb]{0.325\textwidth}
         \centering
         \includegraphics[width=1.0\textwidth,height=1.6in]{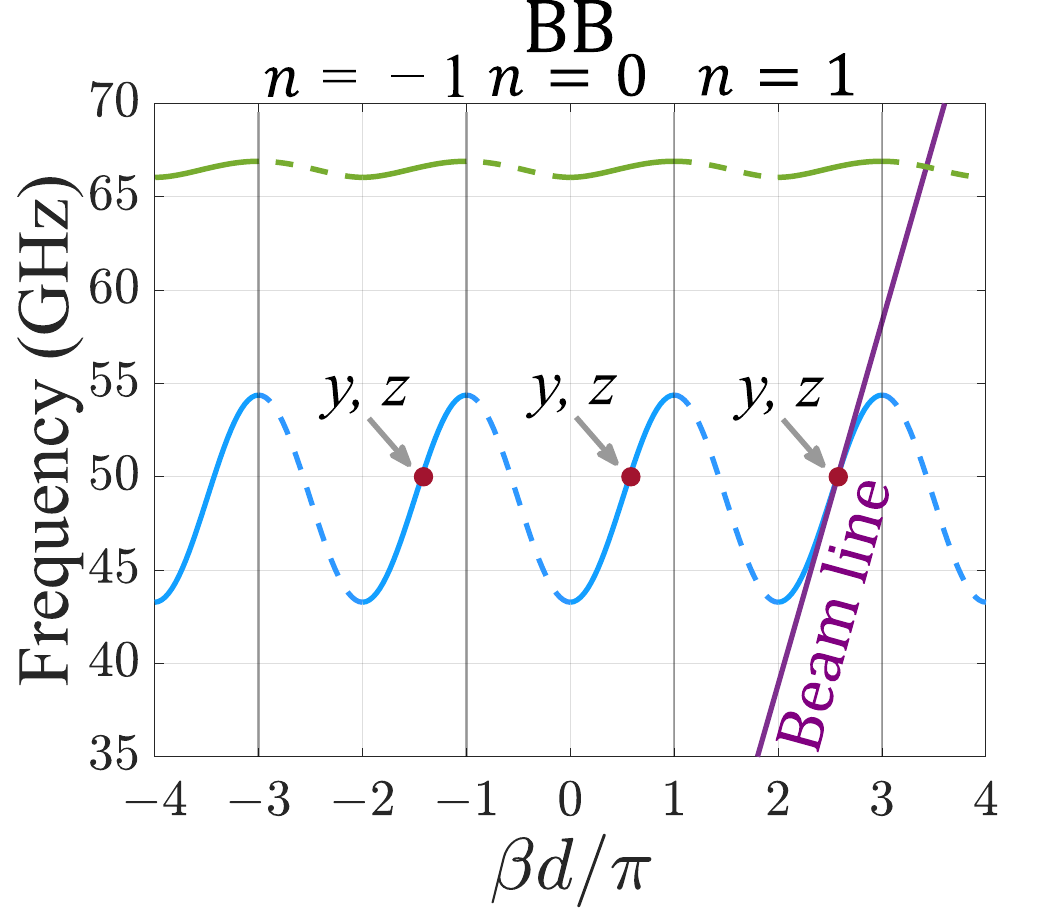}
         \caption{}
         \label{fig:BB_diagram}
     \end{subfigure} 
    \hfill
     \begin{subfigure}[htpb]{0.325\textwidth}
         \centering
         \includegraphics[width=1\textwidth,height=1.6in]{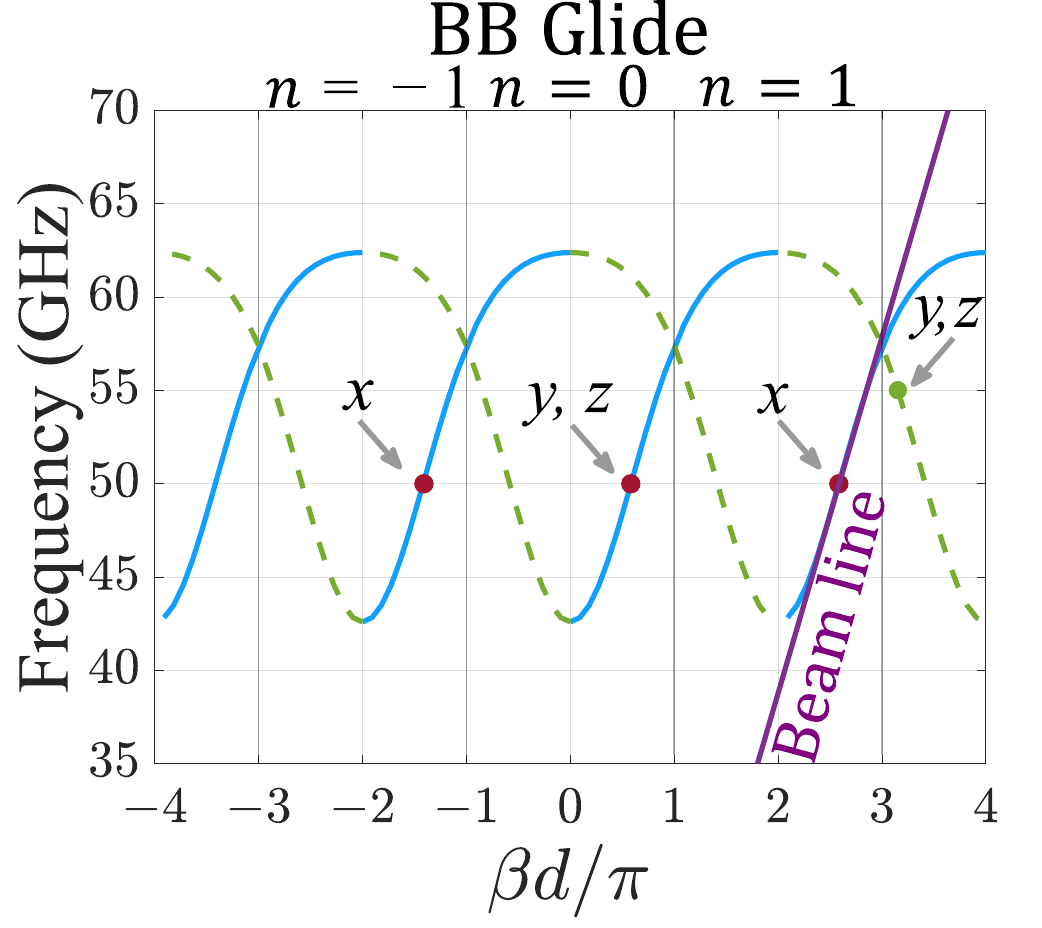}
         \caption{}
         \label{fig:BB_glide_diagram}
     \end{subfigure}
     \hfill
     \begin{subfigure}[htpb]{0.33\textwidth}
         \centering
         \includegraphics[width=1\textwidth,height=1.6in]{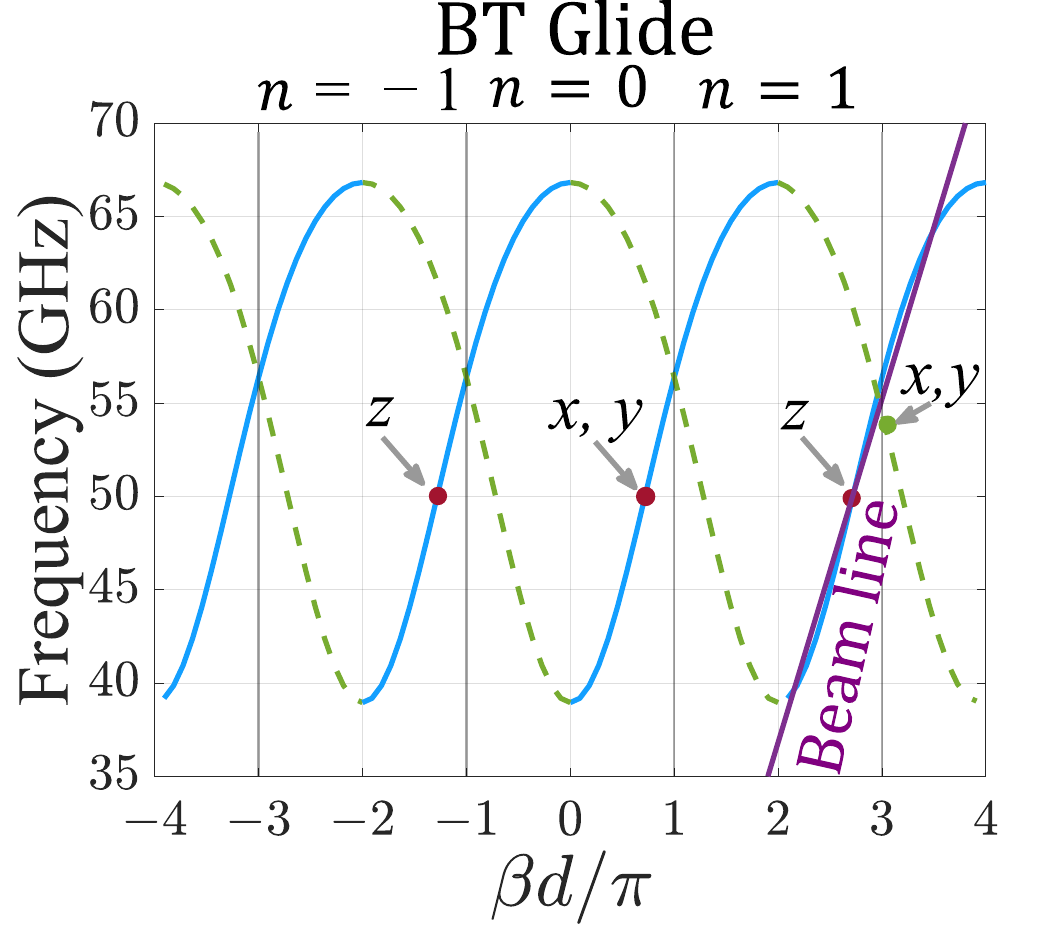}
         \caption{}
         \label{fig:BT_glide_diagram}
     \end{subfigure}
        \caption{Modal dispersion diagrams: (a) BB, (b) BB Glide, (c) BT Glide. The {componets} of the field associated with each Floquet harmonic  {is shown at 50 GHz}. Continuous and dashed lines denote harmonics with positive and negative slope, respectively, and the beam line $\beta_0=\omega/ u_0$ (purple) denotes the electron beam's charge wave with velocity $u_0$. When using glide symmetry in (b) and (c), the two modal branches (green and blue) in (a) merge, and there are no more band edges at $\beta d/\pi = \pm 1, \pm 3,...$. This phenomenon, induced by glide symmetry, provides a large bandwidth for synchronization with an electron beam. {We also show the {field componets} associated to the negative slope branch (dashed green) at $55$ GHz, just below the $\beta d/\pi=3$ intersection point (green dot), for the BB Glide and BT Glide cases. The BT Glide case does not have $E_z$,  indicating no {interaction} of the dashed-green branch with the e-beam.}}
        \label{fig:dispersion_diagram}
\vspace{-0.6cm}
\end{figure*}

In this work, the bandwidth enhancement as a result of the presence of glide symmetry in corrugated waveguides is explored. The glide plane is allocated in two different configurations and the interaction impedance and field {componets} at the first space harmonic are studied and compared to the results pertaining to the double corrugated waveguide without glide symmetry that has been previously published \cite{paoloni15, mineo10, paoloni13, paoloni14}, referred here as "BB" (see Fig. \ref{fig:unit-cells}). By looking at the spatial $\beta$ spectrum, we also establish which modes can be excited in the glide symmetric structures and lead to the conclusion that the newly proposed structure with glide symmetry, here referred to as "BT Glide" (Fig. \ref{fig:unit-cells}) leads to very wide band, dispersionless, operation (Fig. \ref{fig:dispersion_diagram}).


\section{Glide Symmetry and Waveguide Geometries}

\subsection{Glide symmetry}
Glide symmetry is in the two waveguides whose unit cells are shown in Fig. \ref{fig:unit-cells} (b) and (c). A periodic structure is said to possess glide symmetry if it remains invariant under the glide operation $G$ that consists of a translation of half a period $d$, followed by a reflection \cite{Hessel73}. The $G$ operation for the BB Glide structure in Fig. \ref{fig:unit-cells}(b) is defined in Cartesian coordinates as 

\begin{equation}
    G = \begin{cases} 
    x \rightarrow -x \\
    y \rightarrow y \\
    z \rightarrow z + d/2
\end{cases}
\end{equation}
 whereas, for the BT Glide structure in Fig. \ref{fig:unit-cells}(c), the $G$ operation is defined as
 
\begin{equation}
  G =  \begin{cases}
    x \rightarrow -x \\
    y \rightarrow -y \\
    z \rightarrow z + d/2
    \end{cases}
\end{equation}

 In any periodic structure without higher symmetry (like the BB case), the modes exhibit band edges (between a pass and a stop band) at $\beta d/\pi = -1, 0, 1, 2,...$, where the group velocity vanishes. Something different happens for the glide symmetry cases, like the BB Glide and BT Glide geometries. When a structure has glide symmetry,  the upper and lower branches may merge, as it will be clear next, and the group velocity does not vanish at the edge or at the center of the Brillouin zone (BZ): for example, band gaps disappear at $\beta d/\pi= 0, \pm 2, \pm 4,...$ for the cases in Sec. II-IV, whereas band gaps disappear at $\beta d/\pi= \pm 1, \pm 3,...$  for the case studied in the Appendix. The fundamental reason is that, as explained in \cite{Hessel73}, the Bloch-Floquet theorem can be applied to the fundamental cell of length $d/2$ when glide symmetry is present, and not to the full period $d$. In other words, for glide symmetric periodic waveguides the dispersion curves flatten with a spectral period of $4\pi/d$, instead of the usual spectral period of $2\pi/d$ for non-glide symmetric structures.
 
\vspace{-1 mm}

\begin{figure*}
     \centering
     \begin{subfigure}[htpb]{0.32\textwidth}
         \centering
    \includegraphics[width=\textwidth]{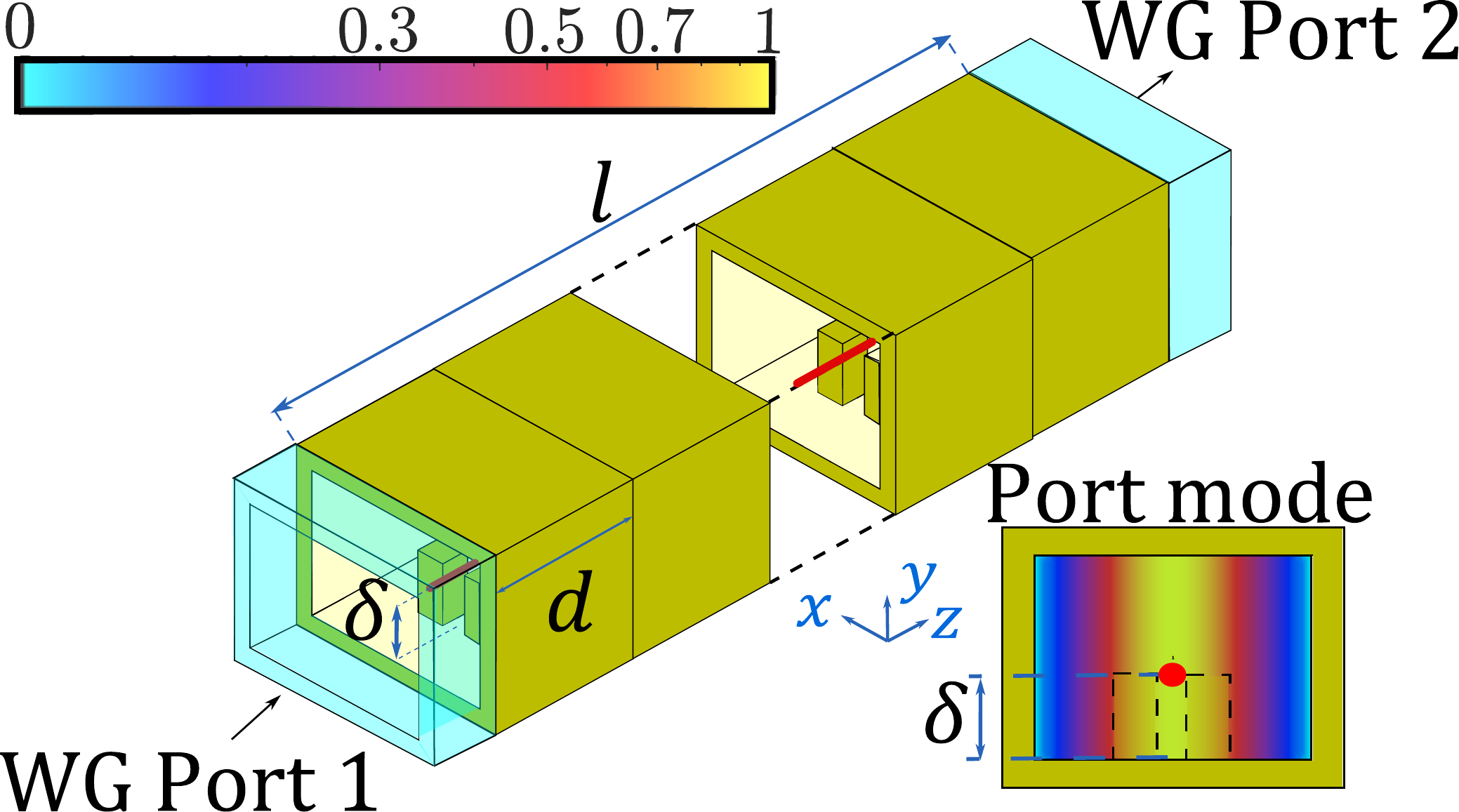}
         \caption{}
         \label{fig:TD_BB_uc}
     \end{subfigure}
     \hfill
     \begin{subfigure}[htpb]{0.32\textwidth}
         \centering
         \includegraphics[width=\textwidth]{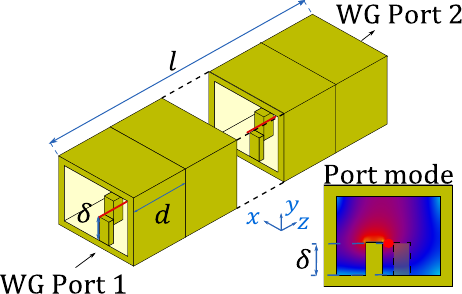}
         \caption{}
         \label{fig:TD_BB_glide_uc}
     \end{subfigure}
     \hfill
     \begin{subfigure}[htpb]{0.3\textwidth}
         \centering
         \includegraphics[width=\textwidth]{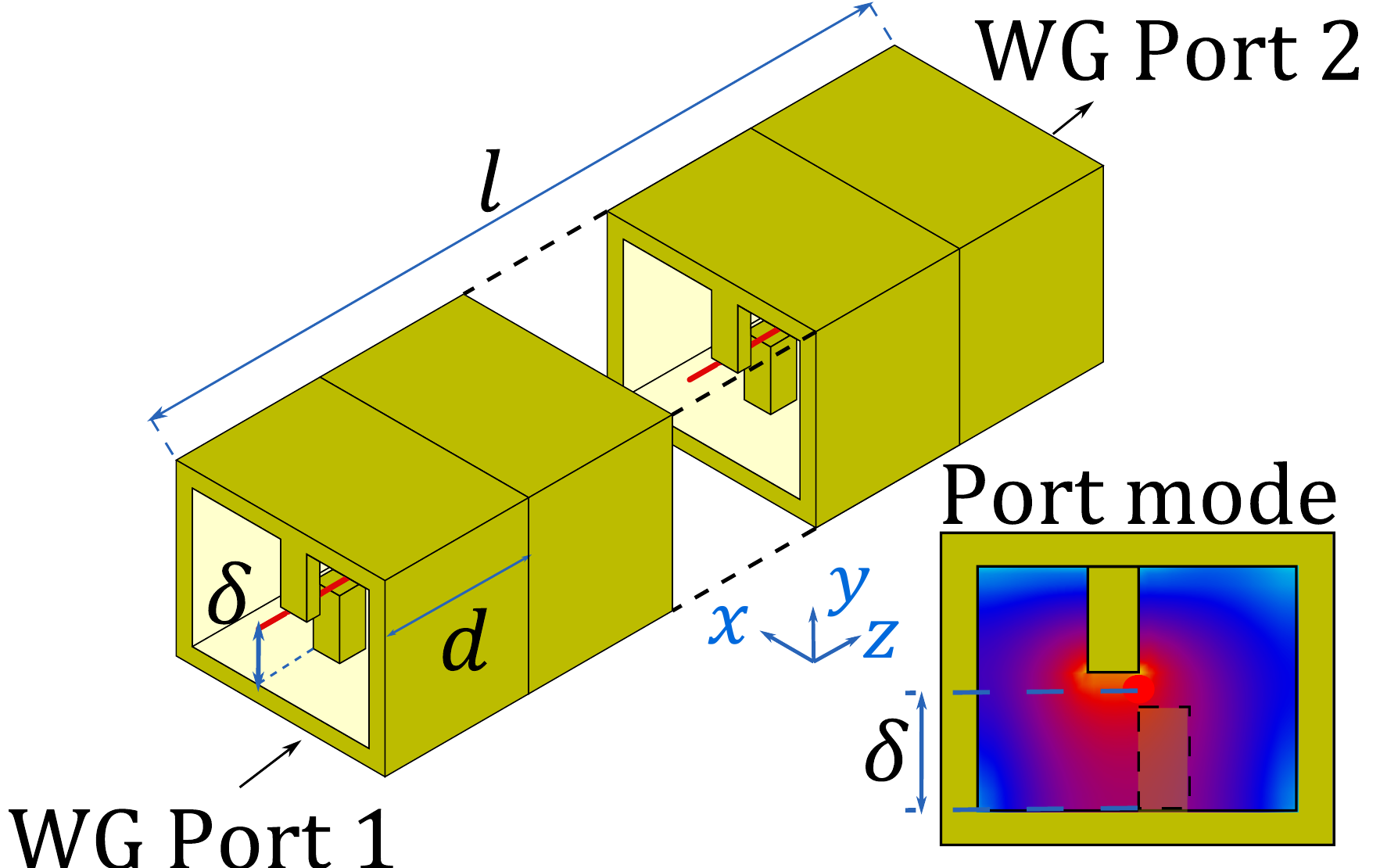}
         \caption{}
         \label{fig:TD_BT_glide_uc}
     \end{subfigure}
        \caption{Simulation setup of the three corrugated waveguides with the corresponding mode excited at the left port: (a) BB case where the light blue waveguide is a small extension used in the computation to have the port slightly away from the pillars;  (b) BB Glide; (c) BT Glide. In these latter two cases, the port includes the section of a metallic pillar.}
        \label{fig:Timedomain_simulation_setup}
        \vspace{-0.5cm}
\end{figure*}

\subsection{Three slow Wave structures}

The so-called double corrugated waveguide (DCW) \cite{paoloni15} has been proposed as a candidate for TWT amplifiers and BWOs \cite{mineo10}. One of its advantages is the wideband modal interaction with the electron beam if the first higher space harmonic is used (i.e., $n = 1$).
In order to improve the wideband behavior of the DCW, the implementation of glide symmetry into the structure is here explored. Three different topologies are studied. First, the classic DCW, which is shown in Fig. \ref{fig:unit-cells}(a), and will be referred to as BB (Bottom - Bottom) in this paper. Then, glide symmetry is implemented considering two different glide planes: first, applying glide symmetry in the $x-z$ plane, as in the unit cell shown Fig.\ref{fig:unit-cells}(b) that will be referred to as BB Glide, which corresponds to the offset(O)-DCW in \cite{Basu2021}; and finally glide symmetry is in both the $x-z$ and $y-z$ planes, as shown in Fig. \ref{fig:unit-cells}(c) and referred to as BT (Bottom - Top) Glide. An interesting case of BB Glide symmetry is further analyzed in the Appendix. The three structures will be analyzed in terms of the dispersion diagram, where the mode excitation and {componets} of each space harmonic will be considered, and then later on looking at the spectrum of the excited modes, and finally in terms of interaction (Pierce) impedance. 


\subsection{Dispersion diagrams}

The dispersion diagram of the modes of each waveguide is computed using the eigenmode solver of the commercial software, CST Microwave Studio assuming the metallic waveguides are made of perfect electric conductor. The unit cells are optimized to have the $n = 1$ harmonic of the electromagnetic mode interact (synchronize) with an electron beam with a charges' velocity of 0.28$c$  for the BB and BB glide cases, and $u_0=0.275 c$ for the BT glide case, being $c$ the speed of light. This leads to the following dimensions (in mm), $a = 2.7$, $b = 2.1$, $a_{pin}$ = 0.4, $h_{pin}$ = 0.9, $d = 2.2$. For the BB and the BB Glide cases, we assume $g = 0.3$ mm, whereas for the BT Glide case we have $c = 0.3$ mm and $g = 0$ mm. The values of {$d$,} $g$ and $c$ are {the same in the three SWS} for comparison purposes. (A further case of BB glide symmetry is analyzed in the Appendix.). The obtained dispersion diagrams are plotted in Fig. \ref{fig:dispersion_diagram}. 

We use the common definition for the BZ, which  implies that the fundamental (0th order) BZ is defined for $-1 < \beta d/\pi <1$. Therefore the $n=1$ Floquet harmonic is defined for  $1 < \beta d/\pi <3$, denoted as first high order BZ. 
For the BB case, the modal dispersion diagram in Fig. \ref{fig:dispersion_diagram}(a) is showing the first two lower modal branches, where a stop band exists between them. The interaction with an electron beam may occur between 45 GHz and 53 GHz, and a fractional bandwidth of 16\% is obtained at 50 GHz which is in line with previously published works with this topology. When glide symmetry comes into the structure, the stop band between the two modes is eliminated leading to a larger bandwidth, for both BB and BT Glide SWSs because of the non dispersive $k-\omega$ relation over a very wide frequency range, especially for the BT Glide case. In the BB Glide case, the diagram in  Fig. \ref{fig:dispersion_diagram}(b) shows that the fractional bandwidth is around 28\%, whereas for the BT Glide SWS, the diagram in Fig. \ref{fig:dispersion_diagram}(c) shows a bandwidth of approximately 50\%, i.e., between 40 GHz to 65 GHz. 

{We performed a simulation for the BB case with period $d'=d/2$ (not shown here for brevity) and observed that the passband width (solid blue branch) increased at the expenses of a larger dispersion that does not provide the possibility to synchronize the EM wave with the e-beam over a wide band.}
It is necessary then to study the {electric field components of the modes} associated to the branches of the dispersion diagram; this information is important to understand what happens when a source is located at one end of the waveguide, as the intersection of the branches of the dispersion diagram at odd values of $n$ may lead to unwanted behavior. 

\begin{figure*}
     \centering
     \begin{subfigure}[htpb]{0.325\textwidth}
         \centering
       \includegraphics[width=1.05\textwidth]{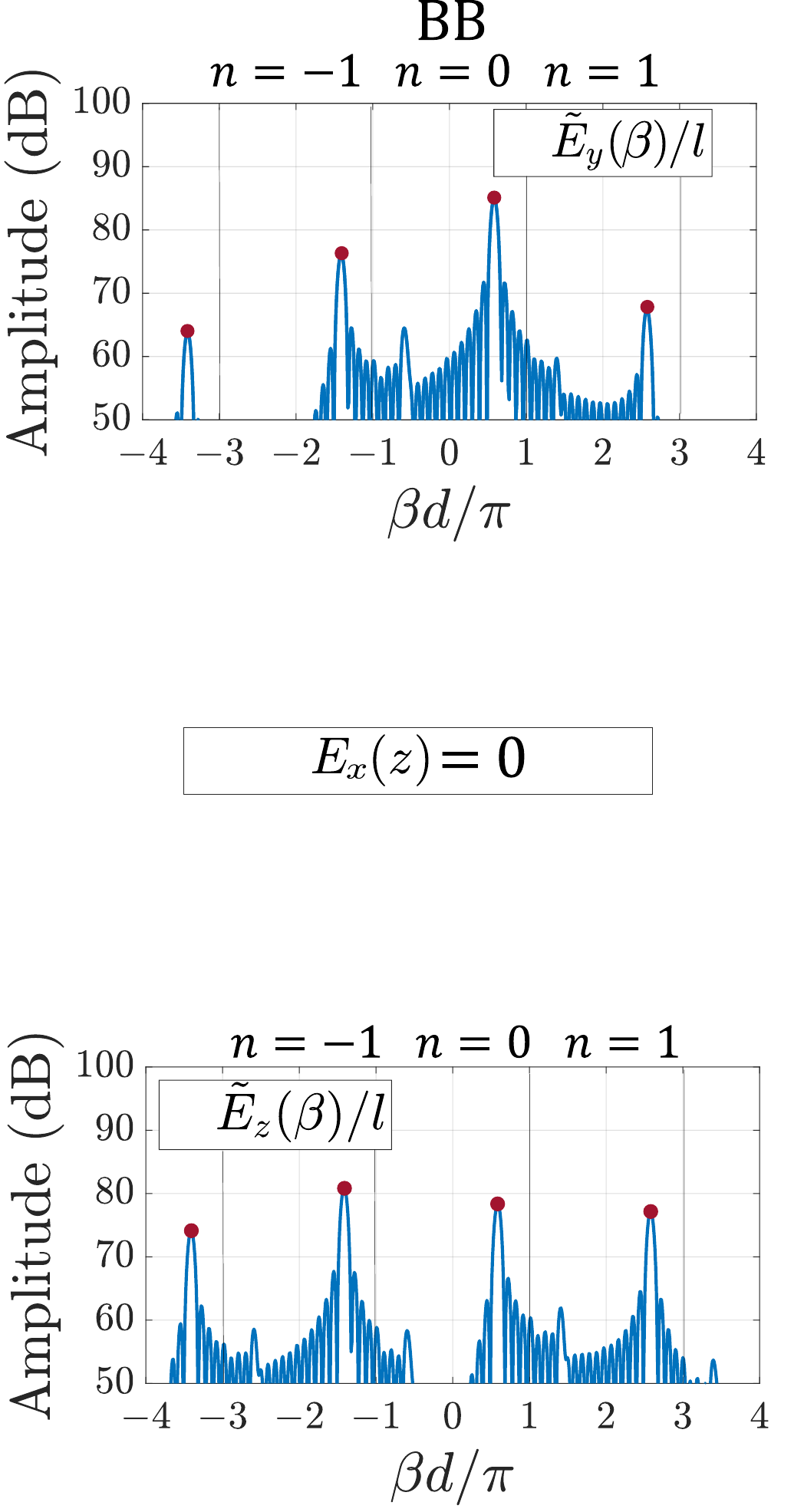}
         \caption{}
         \label{fig:FFT_BB}
     \end{subfigure}
    \hfill
     \begin{subfigure}[htpb]{0.325\textwidth}
         \centering
        \includegraphics[width=1.05\textwidth]{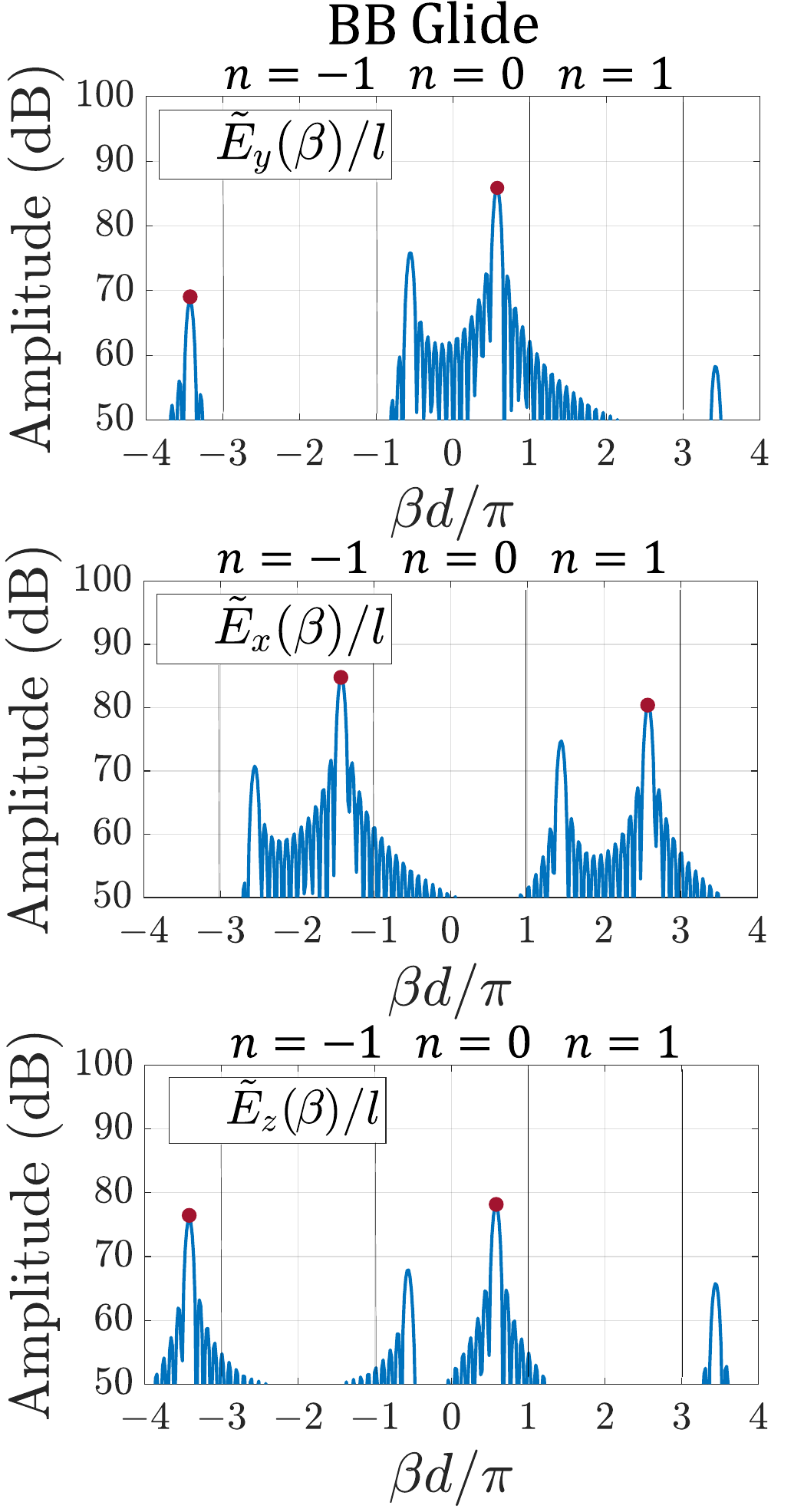}
         \caption{}
         \label{fig:FFT_BB_GLIDE}
     \end{subfigure}
    \hfill
     \begin{subfigure}[htpb]{0.325\textwidth}
         \centering
          \includegraphics[width=1.05\textwidth]{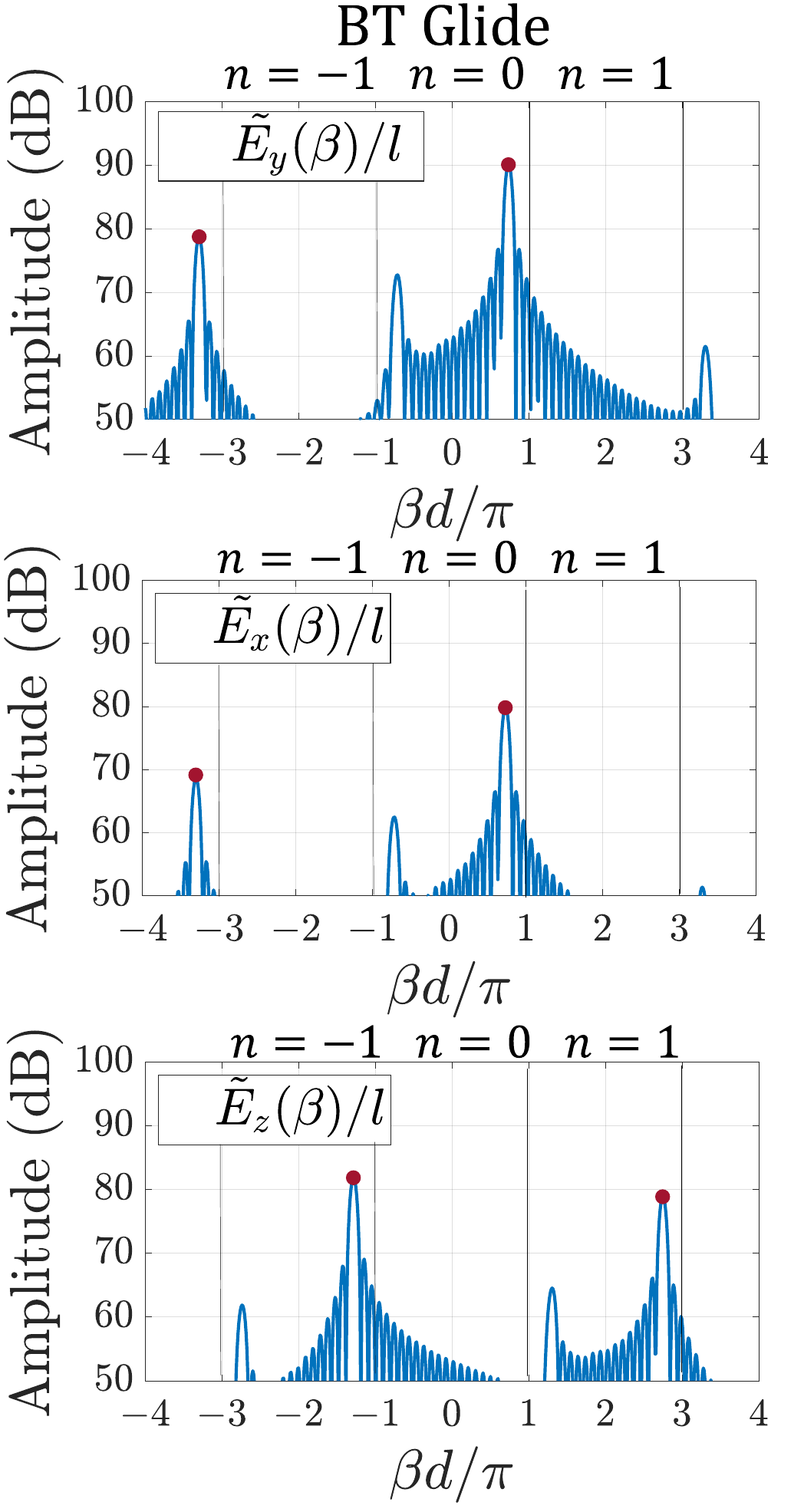}
         \caption{}
         \label{fig:FFT_BT_GLIDE}
     \end{subfigure}
        \caption{Spatial spectrum of the electric field components $E_x$, $E_y$, and $E_z$ at 50 GHz excited by a source at the left end of the SWS: (a) BB case;  (b) BB Glide case; (c) BT Glide case. We are interested in generating a strong $E_z$ component in the $n=1$ space harmonic. The red dots represent the amplitudes related to the branches with positive slope, and the peaks with no dots are related to the waves reflected toward the source from a SWS truncation at the right end.}
        \label{fig:beta_spectrum}
\vspace{-0.5cm}     \end{figure*}

    The solid lines in the dispersion diagrams in Fig. \ref{fig:dispersion_diagram} represent the excited branches by a source at the left end of the waveguide, since this mode has positive group velocity and carry energy along the $+z$ direction. The {components of the electric field} of each harmonic was determined by performing the Fourier analysis of each component of the electric field phasor, $E_x(z)$, $E_y(z)$ and $E_z(z)$, solution of the eigenmode solver. The results show that for the three different geometries, different {componets} of the $n$th harmonic are obtained. For the BB case, all branches have the same {electric field components} but when Glide symmetry is implemented, the branch associated with the $n = \pm 1$ harmonic has different {field components} depending on which symmetry plane is taken into account. In order to have {interaction} with the electron beam, the field must be {polarized} in the direction in which the electrons are traveling \cite{schachter11CH1}; therefore, from this analysis,  the BB Glide is not suitable for wideband interaction at the first space harmonic (i.e., $n=1$), and this will be confirmed later on when the interaction impedance for this structure is computed. Instead, for the BT Glide case, the $n=1$ harmonic of the field is {polarized} in the $z$ direction and hence suitable for {interaction} with an electron beam. It is also important to determine the {field components} of the backward mode in the BB Glide and BT Glide cases, associated with the negative slope branch (dashed green) near the intersection with the positive slope branch (solid blue) at $\beta d/\pi=3$. Indeed, it is around that point that the beam line intersects with the negative slope branch and it could form the onset for a backward wave oscillator. We observe that the field associated with the negative slope branch (dashed green) of the BB Glide case is $z$ {polarized} at the green dot location, whereas the field associated with the negative slope branch (dashed green) of the BT Glide case is {\em not} $z$ {polarized} at the green dot location.

\section{Mode and Spatial Harmonic Excitation}
We investigate what modes are actually excited and what are their {electric field components} in a finite-length SWS excited by Port 1 on the left, as shown in Fig. \ref{fig:Timedomain_simulation_setup}.
Time-domain full-wave simulations have been performed for a perfect conductor waveguide of length $l=20d$, where $d= 2.2$ mm. The field is calculated along a line in the $z$ direction, {at the center of the waveguide in the $y$ direction and at a hight $\delta$}: for the BB and BB Glide $\delta = 0.9$ mm whereas for BT Glide $\delta = b/2 = 1.05$ mm. For each case, we choose waveguide ports as follows. For the BB case, Port 1 excitation corresponds to a conventional $TE_{10}$ mode, Fig. \ref{fig:Timedomain_simulation_setup}(a), and to better match this port mode to the mode of the corrugated waveguide with lower cutoff frequency, an extra waveguide length between the port and the pillar is added. For the BB Glide and BT Glide cases, the port is located in contact with a pillar.  
The phasor of each electric field component  $E_v(z)$, with $v = x,y,z$. is numerically calculated on a scan line  along the $z$ direction (red line) and exported to perform the Fourier transform  

\vspace{-0.3cm}
\begin{gather}
    \tilde{E}_{v}(\beta) = \int_{z=0}^{z=l} E_{v}(z)e^{j\beta z} dz.  \label{eqn: fourier}
\end{gather}

\begin{figure}[htpb]
\begin{subfigure}{.50\textwidth}
  \centering
  \includegraphics[width=0.8\linewidth]{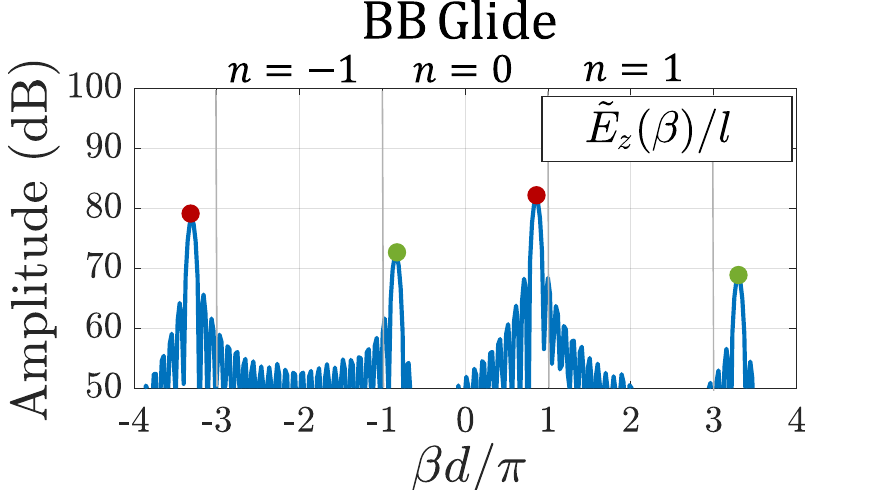}  
  \caption{}
  \label{fig:Spectrum_55GHz_BBGlide}
\end{subfigure}
\begin{subfigure}{.50\textwidth}
  \centering
  \includegraphics[width=0.78\linewidth]{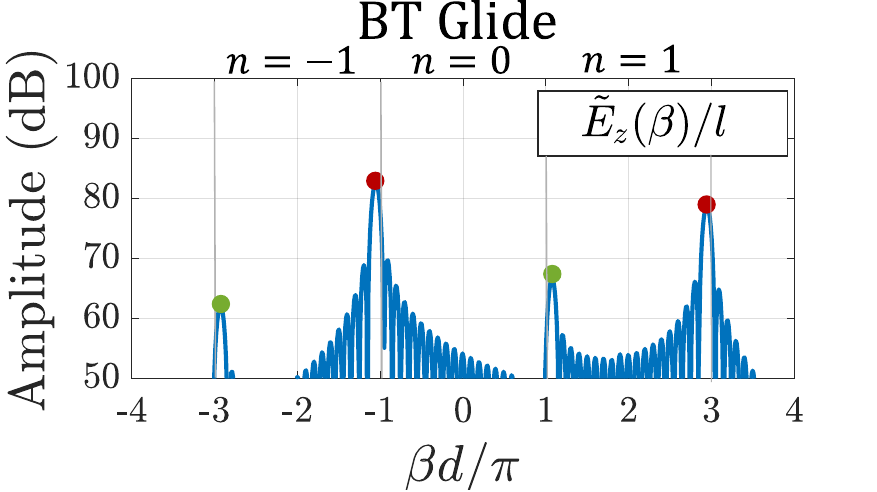}  
  \caption{}
  \label{fig:Spectrum_55GHz_BTGlide}
\end{subfigure}
\caption{{Spatial spectrum of $E_z(z)$ at 55 GHz, i.e., just below the $\beta d/\pi = 3$ intersection point between the two branches in Fig. \ref{fig:dispersion_diagram}. There, the backward wave (green dashed in Fig. \ref{fig:dispersion_diagram})  of the BB Glide case contains the $E_z$ {component} (green dot), whereas the BT Glide case does not have $E_z$ {component} (absence of green dot).}} 
\label{fig:Spectrum_beta3}
\end{figure}
\noindent where $\beta$ is the spectral wavenumber. The spatial spectrum  $\tilde{E}_{v}(\beta)$ at 50 GHz is shown in Fig. \ref{fig:beta_spectrum}, as $20 \log\left[|\tilde{E}_v/l|/(1 \rm{V/m}) \right]$, {assuming all three cases are excited by the same power}. The amplitudes of the red-dotted peaks are related to the mode with positive group velocity, at different spatial harmonics $n$. {In all cases, the harmonic that would synchronize with the electron beam is the $n=1$.}
 In the BB case, we observe $z$ polarized field in the $n=-1$, $n=0$, and $n=1$   spatial harmonics.  {Therefore, the $n=1$ harmonic may be chosen to interact with an electron beam.
 For the BB Glide in Fig. \ref{fig:beta_spectrum}(b), in the fundamental Brillouin zone region from $-1<\beta d/\pi<1$ (i.e, $n=0$), the field only has components related to $y$ and $z$, but when looking at the first harmonic region, $n = 1$, the electric field is $x$ polarized, and there is no $z$ polarized field, {hence preventing the {interaction} with an electron beam}. Furthermore, {in the studied  BB Glide case, also the synchronization in the fundamental Brillouin zone (i.e., $n=0$)  is not possible because the mode there} exhibits a fast-wave behavior since the phase velocity is greater than the speed of light in most of the frequency band.
 On the other hand, for the BT Glide case, the peaks of the amplitude of the field are $y$ and $x$ polarized in the fundamental BZ (i.e., $n=0$) and $z$ polarized for the $n = 1$ spatial harmonic. Therefore, {{interaction} with an electron beam is possible, and} the BT Glide SWS may be chosen as a SWS for a TWT operating with the $n=1$ spectral region. 
  This analysis allows us to evaluate which SWS is suitable for a TWT amplifier application because we need a strong $E_{z,n}$ spatial harmonic (see next section) in the spectral region $n=1$. We conclude that if wide band interaction is required, the particular case of BB Glide SWS here studied cannot be used to work in the $n = 1$ spectral region ($E_{z,1}$ is negligible in Fig. \ref{fig:beta_spectrum}); however this structure could be used in a TWT if the RF wave is slow enough and $d<<\lambda$ using a large interaction impedance at the fundamental harmonic since $E_{z,0}$ is strong. For the BT Glide case, it is possible to achieve a very wide band synchronization in the first ($n=1$) harmonic region to design wideband power amplifiers. 
 An important observation from the results of this section is that a Port 1 excitation would not excite a mode with negative group velocity shown in Fig. \ref{fig:dispersion_diagram} with dashed curves. A Port 1 excitation at 50 GHz would excite only a mode {pertaining to the branches} with positive group velocity. A backward mode associated with dashed curves (harmonics with negative slope) can be excited only by a wave reflected by Port 2. The higher order mode for the BB case (green line at higher frequency in Fig. \ref{fig:dispersion_diagram}), could synchronize with the electron beam generating self-sustained oscillations. However, the higher mode for the BB case (green curve, in Fig. \ref{fig:dispersion_diagram}) is not present anymore for the BB Glide and BT Glide cases because it merged with the lower mode (Fig. \ref{fig:dispersion_diagram}), thanks to glide symmetry. Importantly, as demonstrated in Fig. \ref{fig:beta_spectrum}, {for the BB Glide and BT Glide cases} only the positive-group velocity mode (solid blue, in Fig. \ref{fig:dispersion_diagram}) can be excited by Port 1. {Keeping in mind the dashed-green 
 negative-slope branches in Fig. \ref{fig:dispersion_diagram}, Fig. \ref{fig:beta_spectrum} also confirms that at 50 GHz  the weak backward mode is excited by the reflection at Port 2. From Fig. \ref{fig:beta_spectrum} we also observe that there is a small $z$-polarized field in the $n=2$ harmonic ($3 < \beta d/\pi < 5$) of the reflected wave for the BB Glide case, whereas the $z$ {component} is not present in the $n=2$ harmonic for the BT Glide case.}

{Another important result is that at the intersection between the beam line and the negative-slope branch (dashed green) in Fig. \ref{fig:dispersion_diagram}, the electromagnetic wave does not have $z$ {component} in the BT Glide case. This is further confirmed by the spatial spectrum of the $E_z(z)$ field in Fig. \ref{fig:Spectrum_beta3} of the BT Glide case. Indeed, the spatial spectra in Fig. \ref{fig:Spectrum_beta3} correspond to the green dot points in Fig. \ref{fig:dispersion_diagram} for the BB Glide and BT Glide cases at $f$= 55 GHz, where the normalized wavenumber $\beta d/\pi$ is slightly larger than 3. It is clear that the field of the BB Glide case exhibits a $z$ {component} (green dot at $\beta d/\pi > 3$), whereas the field of the BT Glide case is not $z$ polarized (absence of green dot at $\beta d/\pi > 3$)}. {In summary, in a TWT device, the electron beam would not synchronize with a backward electromagnetic wave if BT Glide geometry is considered, since no $E_{z}$ component can be seen in the spatial spectral region between $3 < \beta d/\pi < 4$,} eliminating the distributed feedback mechanism {risk} for a BWO. 



\section{Interaction impedance}

In a TWT, the coupling between the RF wave and the electron beam is described by the interaction (or Pierce) impedance, $Z_n$, which is the ability of said wave to exchange energy with the synchronized electron beam. A large interaction impedance represents the ability of the SWS to guide modes with a strong $E_z$ field at the location where the electron beam would flow \cite{pierce-twt50}. The interaction impedance is defined as a measure of the longitudinal electric field acting on the charges for a given total power flow in the structure \cite{schachter11CH1}, that is 
\begin{equation}
    Z_n = \frac{|E_{z,n}|^2}{2\beta_n^2P}   \qquad n = 0, \pm 1, \pm2,...     \label{eqn:interaction_impedance}
\end{equation}

\noindent where $\beta_n = \beta_0 + 2\pi n/d$, is the wavenumber of the $n$th Floquet harmonic, $P$ is the mode time-average electromagnetic power flow \cite{pierce-twt50}. Finally, $E_{z,n}$ is the phasor representing the longitudinal component of the $n$th harmonic of the electric field $E_{z}(z)$ in the periodic SWS, obtained from the Floquet-Bloch series expansion $E_{z}(z) = \sum_{n=-\infty}^{\infty}E_{z,n}e^{-j\beta_n z}$. The spatial harmonic strength $E_{z,n}$ is found as

\begin{equation}
    E_{z,n} = \frac{1}{d}\int_{0}^{d} E_{z}(z)e^{j\beta_n z} dz.  \label{eqn: fourier2}
\end{equation}
{The phasor of the field $E_z(z)$ is calculated using eigenmode simulations imposing phase-shift-periodic boundary conditions on a unit cell of the SWS, at the same location discussed in the previous section}.  
\begin{figure}[htpb]
\centering
\includegraphics[width=0.8\columnwidth]{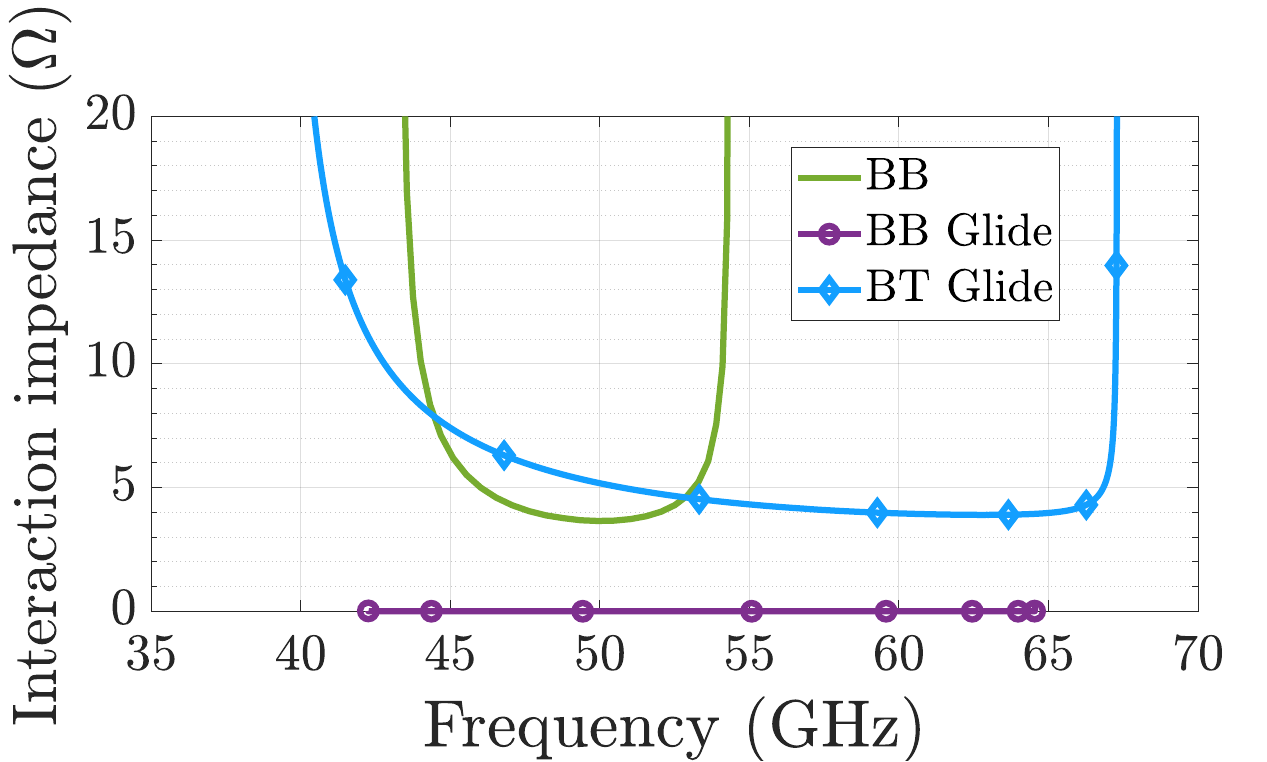}
\caption{Interaction impedance for the $n = 1$ Floquet harmonic with positive slope for the three SWSs. The BT Glide SWS exhibits a very large operating frequency band because of glide symmetry. The BB Glide case has a vanishing $E_z$ component in the $n=1$ harmonic, hence  a vanishing impedance. The BB case has a narrower bandwidth because the structure does not have glide symmetry.}
\label{fig:Interaction_impedance}
\end{figure}

\begin{figure*}[htpb]
     \centering
     \begin{subfigure}[htpb]{0.328\textwidth}
         \centering
         \includegraphics[width=1.05\textwidth]{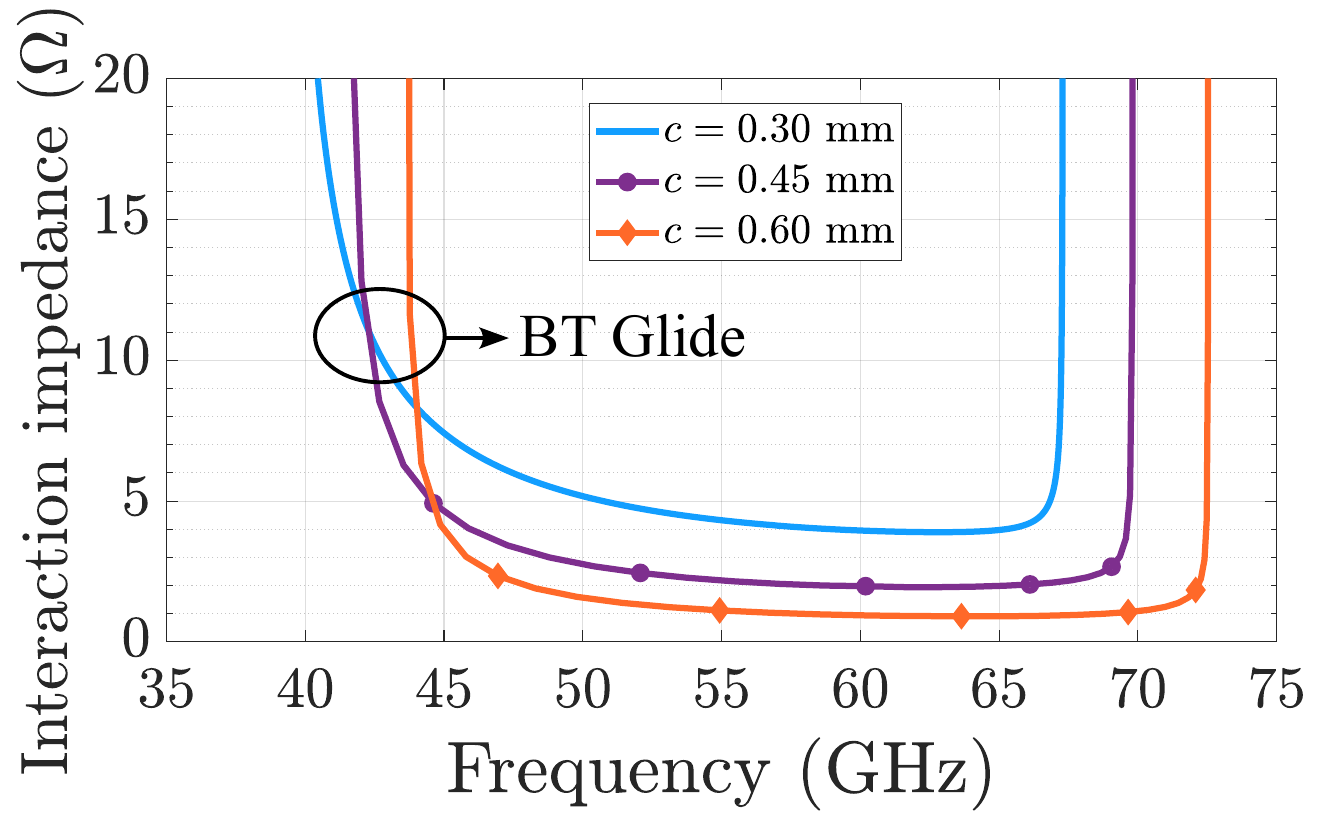}
         \caption{}
         \label{fig:c_parametric}
     \end{subfigure}
    \hfill
     \begin{subfigure}[htpb]{0.328\textwidth}
         \centering
         \includegraphics[width=1.05\textwidth]{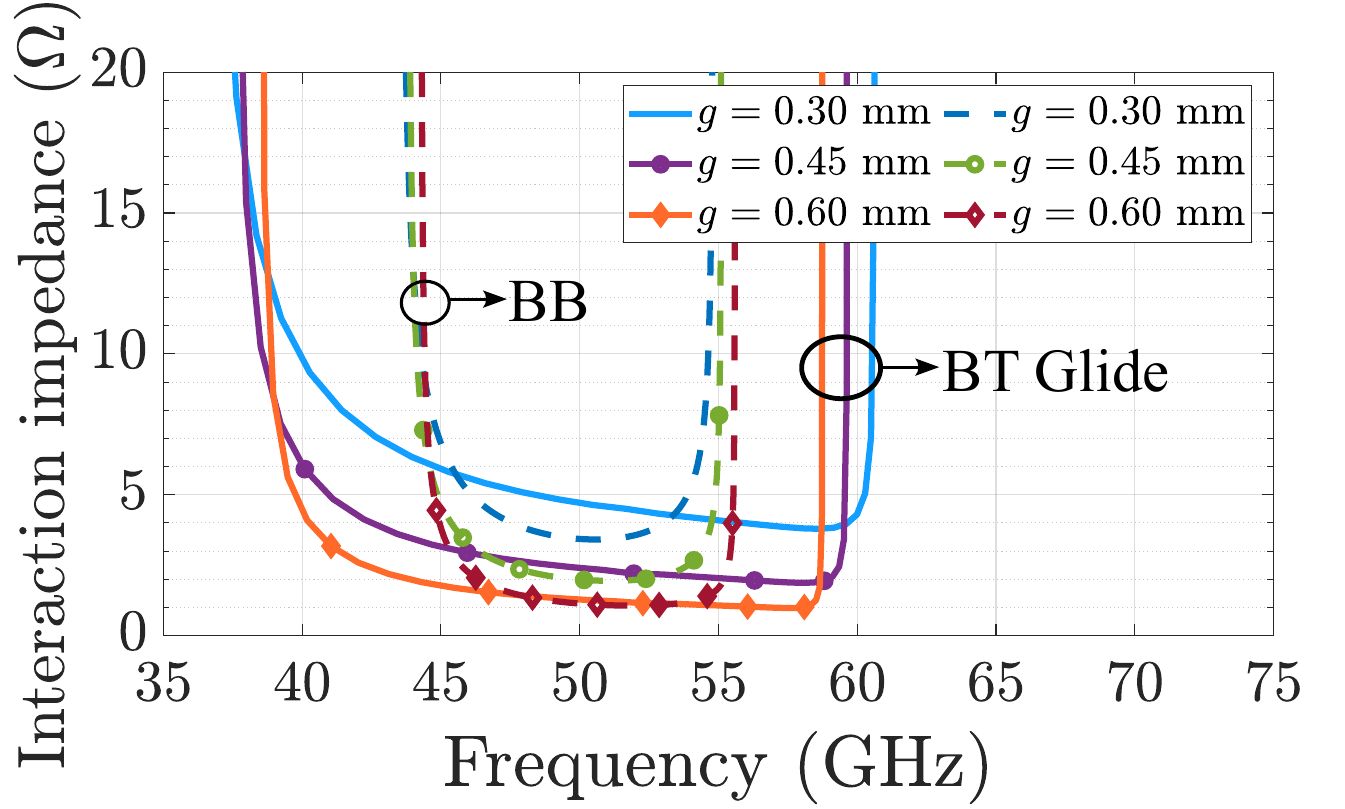}
         \caption{}
         \label{fig:g_parametric}
     \end{subfigure}
     \hfill
     \begin{subfigure}[htpb]{0.328\textwidth}
         \centering
         \includegraphics[width=1.05\textwidth]{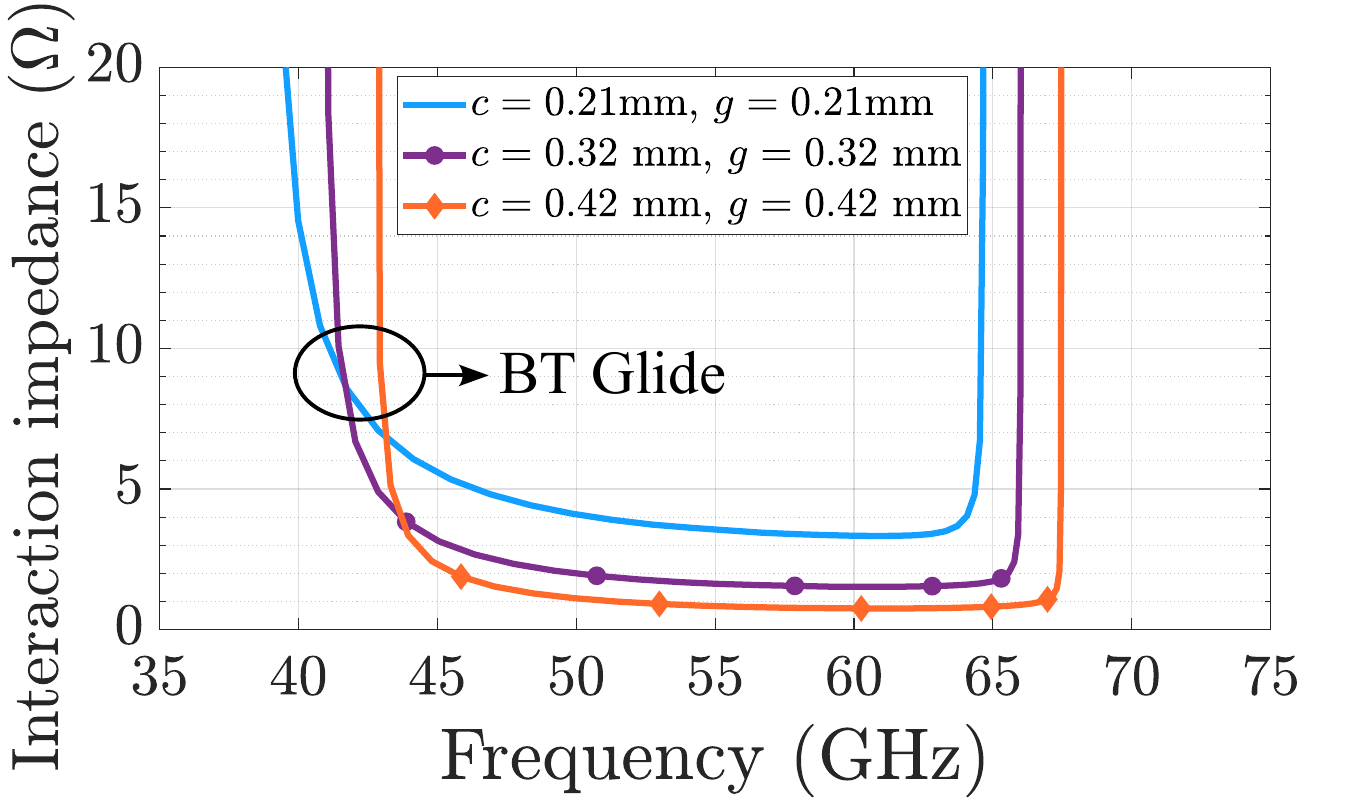}
         \caption{}
         \label{fig:c_g_parametric}
     \end{subfigure}
        \caption{Parametric study of the interaction impedance $Z_1$, associated to the dispersion branch with positive slope, for different spacings between the pillars for the BT Glide case: (a) Different values of $c$ while $g = 0$ mm. (b) Different values of separation $g$ while $c = 0$ mm. The impedance values are comparable with those of the BB geometry but BT Glide shows much wider bandwidth. (c) Different values of pillar separation with increments in $c$ and $g$.}
        \label{fig:interaction_impedance_beam_channel}
\end{figure*}


\noindent 

In this paper, we are particularly interested in the branch with a positive slope of the $n = 1$ harmonic, which is the one that would interact with the electron beam, aiming for wideband velocity synchronization, and where the group velocity is positive and the SWS can be used as an amplifier. The interaction impedance versus frequency for the positive-slope branch of the modes in the three geometries is plotted in Fig. \ref{fig:Interaction_impedance}. We can see that, as expected, for the BB and BT Glide cases we have a nonzero interaction impedance, but for the BB Glide case, the impedance is vanishing because of the null of the longitudinal $E$-field component of the $n=1$ harmonic (see also the spectrum in Fig. \ref{fig:beta_spectrum}) as observed also in \cite{Basu2021} for the $n=1$ harmonic component of the lower-band mode (see also the distinct case in Appendix where a more complex mode structure exhibits a non-vanishing impedance for the upper mode with positive slope). 

In the BB and in the BT Glide cases, the results show that $Z_n \geq 4$ $\Omega$ over the relative frequency bands; this value can be optimized by changing the dimensions of the corrugated waveguide such as pins height, beam position, beam channel width, and so on. What is very clear is that the use of glide symmetry dramatically increases the bandwidth to more-or-less a relative value of 34\% without causing any reduction in the interaction impedance compared to the BB case.

{One constraint that this family of SWS has is the need to place the beam close to the metal corrugations to maximize the interaction impedance as shown in \cite{paoloni15}. This may result in the interception of the traveling electrons. A parametric study on the relationship between the interaction impedance and the spacing available in between corrugations to fit the electron beam is performed}. The variable parameters are $c$ and $g$, shown in Fig. \ref{fig:unit-cells}. 
First, in the original BT Glide case, where $g = 0$ mm, the spacing is controlled by $c$, leading to the results in Fig. \ref{fig:interaction_impedance_beam_channel}(a), with a range from $c = 0.3$ mm to $c = 0.6$ mm, with a step of $0.15$ mm. A similar procedure is performed for the results in Fig. \ref{fig:interaction_impedance_beam_channel}(b) by setting $c = 0$ mm so that the spacing is controlled by $g$ only, and the results are compared with those of the BB case because in such a case the distance between pillars depends exclusively on $g$. Finally, for the results in Fig. \ref{fig:interaction_impedance_beam_channel}(c), both $c$ and $g$ are set to have a spacing from $0.3$ mm to $0.6$ mm in the cross-sectional diagonal between the corrugation pillars.  
From the latter results, we can see that the interaction impedance decreases as the separation between pillars increases, with the worst case being  $Z_n \approx 1$ $\Omega $ when the distance is $0.6$ mm. However, the comparison shown in Fig. \ref{fig:interaction_impedance_beam_channel}(b) shows that the decrease in the interaction impedance with increasing spacing between corrugations is comparable with that of the BB geometry and that the wideband benefits of glide symmetry are still present: this means that BB and BT Glide will have a similar (or equal) value of impedance, but in presence of glide symmetry the bandwidth obtained is much larger. The precise selection of the geometric values will depend on the beam velocity, operating frequency band, required beam current, etc; the BT Glide and BB geometries offer several degrees of freedom in the design, including the period $d$ to make the dispersion diagram overlapping the beam line over a wide frequency band. 

\section{Conclusion}

 We have presented a study of the interaction impedance and mode excitation in three double corrugated rectangular waveguides, where one of these, the BT Glide structure, is a novel structure that uses glide symmetry in both the $x-z$ and the $y-z$ planes. The presented structure is suitable to operate around 50 GHz in the first (i.e., $n=1$) space harmonic, where the interaction impedance is greater than 4 $\Omega $ while the fractional bandwidth is approximately 50 \%, which is significantly greater than the bandwidth exhibited by structures without glide symmetry or where the glide symmetry is only applied over one plane. Furthermore, it was shown that depending on the plane in which the glide operator is applied, the excited mode can have a stronger or weaker longitudinal {componet}. This is critical if wideband interaction is required at a given Floquet harmonic. {We have also found that the backward wave of the BT Glide case does not exhibit a $z$ {componet} of the $n=2$ spatial harmonic, whereas such {componet} exists in the $n=2$ harmonic of the backward wave of the BB Glide case. This is the harmonic of the backward wave that would synchronize with the electron beam near the $\beta d/\pi=3$ point leading to a backward wave oscillator unwanted behavior. }
 
 Additionally, a parametric study of the interaction impedance as a function of the distance between corrugations (i.e., pillars) was performed. The findings suggest that employing glide symmetry improves the bandwidth without compromising the interaction impedance, compared to the related non-glide BB corrugated waveguide.
 {The BT Glide slow wave structure discussed in this paper can be scaled to be operative in higher frequency ranges of millimeter waves or THz bands while maintaining its performance in terms of interaction impedance, wide band synchronization, and absence of $z$ {componet} in the backward wave to eliminate the possibility of {interaction} of the backward wave with the electron beam.}



\section*{Acknowledgment}


This material is based upon work partly supported by the Air Force Office of Scientific Research Multidisciplinary Research Program of the University Research Initiative (MURI) under grant number FA9550-20-1-0409 administered through the University of New Mexico. The work was partially funded by the Spanish Government under the grant PID2019-107688RB-C21 from MCIN/AEI/10.13039/501100011033 and in part by the COST (European Cooperation in Science and Technology)
through COST Action Symat under Grant CA18223. Eva Rajo Iglesias, Filippo Capolino and Nelson Castro are with the Department of Signal Theory and Communications, University Carlos III of Madrid, Leganés, 28911, Spain (email: ncastro@pa.uc3m.es). Miguel Saavedra-Melo and Filippo Capolino are with the Department of Electrical Engineering and Computer Science, University of California, Irvine, CA 92697 USA (e-mail: masaave3@uci.edu). FC Thanks the Universidad Carlos III de Madrid for receiving a Chair of Excellence that made this research and collaboration possible. The authors thank DS SIMULIA for providing CST Studio Suite. 
\vspace{-3 mm}
\appendix 
\section*{A different BB Glide SWS} \label{Appendix}

In some cases, the BB Glide geometry in Fig. \ref{fig:unit-cells}(b) leads to interesting dispersion diagrams and useful values of the  interaction impedance associated to the $n = 1$ Floquet harmonic modal branch with positive slope. Let us consider for example a DCW with BB Glide symmetry as in Fig. \ref{fig:unit-cells}(b), with unit cell  dimensions (in mm): $a = 1.8$, $b = 0.45$, $a_{pin}$ = 0.09, $h_{pin}$ = 0.327, $d$ = 0.505 and $g$ = 0.17. Some of these dimensions are the same as those in  \cite{Basu2021}.
The dispersion diagram of the modes in the waveguide is shown in Fig. \ref{fig:BB_Glide_Paoloni}(a). It is rather different from what is presented in Section II:  There are two branches with positive slope in the region $2< \beta d/\pi<3$, the solid green and the solid blue (see Fig. \ref{fig:dispersion_diagram}(b) for comparison). The upper mode in solid green has a positive slope in the region $2 <\beta d/\pi < 2.8$, meaning that this modal branch can be used for synchronization with the electron beam to design a TWT amplifier. The interaction impedance is calculated using (\ref{eqn:interaction_impedance}) and plotted as a function of frequency in Fig. \ref{fig:BB_Glide_Paoloni}(b), for both the  solid-blue lower and solid-green upper modes:  the solid blue mode has a vanishing interaction impedance, exhibiting a behavior similar to the BB Glide case shown in Fig. \ref{fig:Interaction_impedance}. Instead, the interaction impedance plotted in the solid green related to the upper solid-green branch in Fig. \ref{fig:BB_Glide_Paoloni}(a), shows that this mode is suitable for the interaction, having a non-zero interaction impedance (an analogous conclusion was reached in \cite{Basu2021,Patent}). However, even though the interaction impedance is non vanishing for the $n=1$ harmonic component of the higher-band mode, the extended bandwidth as a consequence of the glide symmetry is not present. Indeed, interesting observations are made when we compare the dispersion diagrams in Fig. \ref{fig:dispersion_diagram}(b)  and in Fig. \ref{fig:BB_Glide_Paoloni}(a), both relative to BB Glide SWSs. 
The dispersion diagram in Fig. \ref{fig:dispersion_diagram}(b) shows: (i) a very wide band resulting from the disappearing of the bandgap at $\beta d/\pi=3$, which is in the middle of the branch of positive slope, and (ii) there are no other modes in the same frequency range, besides the presence of the mode with negative slope.
The dispersion diagram of the solid-green upper mode in Fig. \ref{fig:BB_Glide_Paoloni}(a) shows: (i) a much smaller bandwidth, despite the disappearing of the bandgap at $\beta d/\pi=3$; (ii) the solid-blue lower mode with positive slope over a wide frequency band, which has a vanishing interaction impedance, still shows the disappearing of the bandgap at $\beta d/\pi=3$, because of glide symmetry; (iii) both modes however experience the closure of the bandgap at $\beta d/\pi=3$ close to their upper band edge; (iv) there are two modes with positive slope, besides the two modes with negative slope. In other words, the dispersion diagram shown in  Fig. \ref{fig:BB_Glide_Paoloni}(a) is complex and results from few mode interactions. One can also recognize that both dispersion curves have a minimum (either at $\beta d/\pi=0$ or at $\beta d/\pi=2$) between two maxima, which is a condition related to the split band edge (SBE) discussed in \cite{Figotin07SlowWaveRes,Chabanov08StronResTra} and that may lead also to the presence of a degenerate band edge (DBE) \cite{Figotin07SlowWaveRes,Othman17ExpDem,Zheng22SymthMeasDBE} when properly selecting the dimensions. An application of the DBE in linear beam devices is the "multimode synchronization regime" for low threshold BWOs \cite{Othman16LowStarting, Zuboraj17PropDBE}.

\begin{figure}[htpb]
\begin{subfigure}{.45\textwidth}
  \centering
  \includegraphics[width=0.8\linewidth]{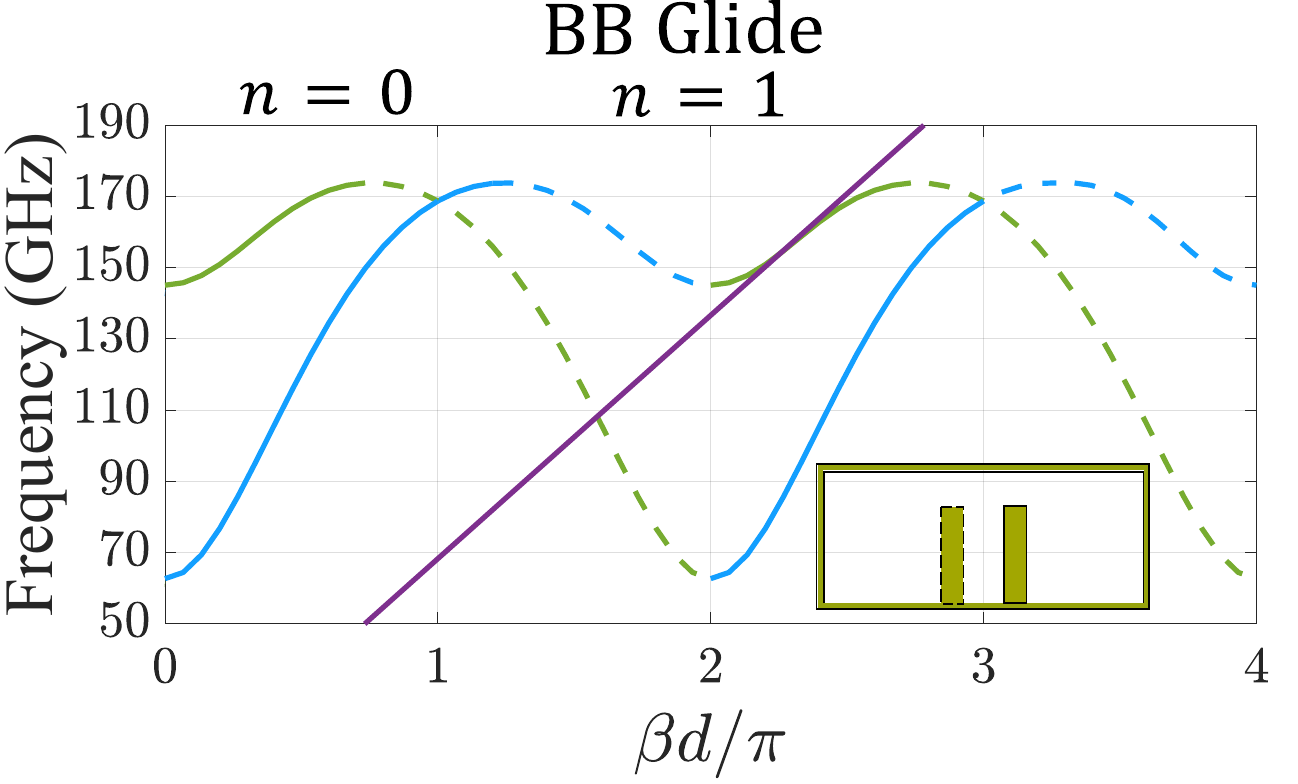}  
  \caption{}
  \label{fig:Z_n_BB_Glide_Paoloni}
\end{subfigure}
\begin{subfigure}{.45\textwidth}
  \centering
\includegraphics[width=0.82\linewidth]{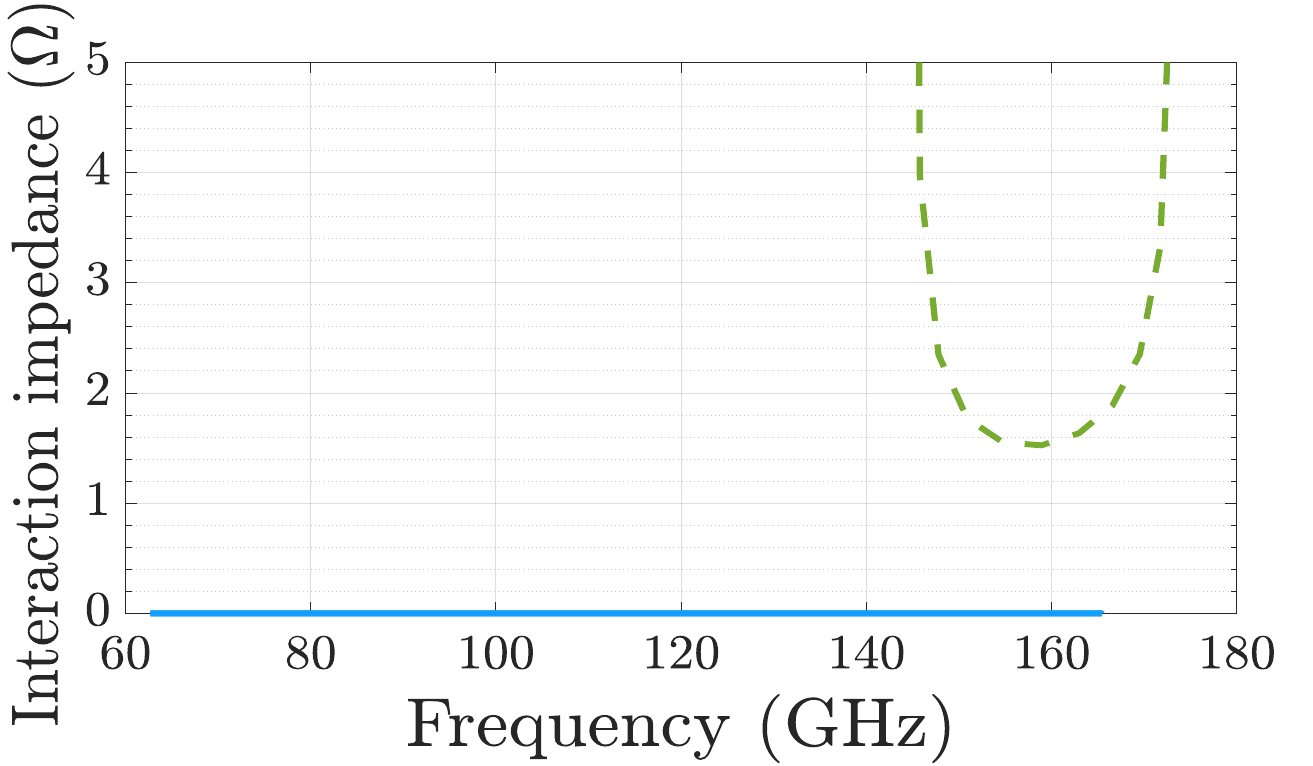}  
  \caption{}
  \label{fig:Z_n_BB_Glide_Paoloni}
\end{subfigure}
\caption{ Results of the BB Glide DCW with dimensions slightly different from the ones in Secs.II-IV.(a) Dispersion diagram, with the beam line plotted for a voltage of 13.5 kV (0.23$c$) for synchronization with the mode in solid green between 150 GHz and 165 GHz. (b) Interaction impedance for the two $n=1$ modal branches with positive slope, in the region $2< \beta d/\pi<3$.} 
\label{fig:BB_Glide_Paoloni}
\vspace{-2mm}
\end{figure}


\bibliographystyle{IEEEtran}
\bibliography{CorrugatedWaveguides.bib}

\begin{thebibliography}{10}
\providecommand{\url}[1]{#1}
\csname url@samestyle\endcsname
\providecommand{\newblock}{\relax}
\providecommand{\bibinfo}[2]{#2}
\providecommand{\BIBentrySTDinterwordspacing}{\spaceskip=0pt\relax}
\providecommand{\BIBentryALTinterwordstretchfactor}{4}
\providecommand{\BIBentryALTinterwordspacing}{\spaceskip=\fontdimen2\font plus
\BIBentryALTinterwordstretchfactor\fontdimen3\font minus
  \fontdimen4\font\relax}
\providecommand{\BIBforeignlanguage}[2]{{%
\expandafter\ifx\csname l@#1\endcsname\relax
\typeout{** WARNING: IEEEtran.bst: No hyphenation pattern has been}%
\typeout{** loaded for the language `#1'. Using the pattern for}%
\typeout{** the default language instead.}%
\else
\language=\csname l@#1\endcsname
\fi
#2}}
\providecommand{\BIBdecl}{\relax}
\BIBdecl

\bibitem{pierce-twt50}
J.~R. Pierce, ``Traveling-wave tubes,'' \emph{The Bell Syst. tech. j}, vol.~29,
  no.~1, pp. 1--59, 1950.

\bibitem{schachter11CH1}
L.~Schächter, \emph{Beam-Wave Interaction in Periodic and Quasi-Periodic
  Structures}.\hskip 1em plus 0.5em minus 0.4em\relax Springer-Verlag Berlin
  Heidelberg, 2011, ch.~1, pp. 10--15.

\bibitem{paoloni21}
C.~Paoloni, D.~Gamzina, R.~Letizia, Y.~Zheng, and N.~C. Luhmann~Jr,
  ``Millimeter wave traveling wave tubes for the 21st century,'' \emph{J.
  Electromagn. Waves Appl}, vol.~35, no.~5, pp. 567--603, 2021.

\bibitem{Booske2008}
J.~H. Booske, ``Plasma physics and related challenges of
  millimeter-wave-to-terahertz and high power microwave generation,''
  \emph{Phys. Plasmas}, vol.~15, no.~5, p. 055502, 2008.

\bibitem{Sengele2009}
S.~Sengele, H.~Jiang, J.~H. Booske, C.~L. Kory, D.~W. van~der Weide, and R.~L.
  Ives, ``Microfabrication and characterization of a selectively metallized
  {W-Band} meander-line {TWT} circuit,'' \emph{IEEE Trans. Electron Devices},
  vol.~56, no.~5, pp. 730--737, 2009.

\bibitem{booke11}
J.~H. Booske, R.~J. Dobbs, C.~D. Joye, C.~L. Kory, G.~R. Neil, G.-S. Park,
  J.~Park, and R.~J. Temkin, ``Vacuum electronic high power terahertz
  sources,'' \emph{IEEE Trans. Terahertz Sci. Technol.}, vol.~1, no.~1, pp.
  54--75, 2011.

\bibitem{Armstrong2012}
C.~M. Armstrong, ``The truth about terahertz,'' \emph{IEEE Spectrum}, vol.~49,
  no.~9, pp. 36--41, 2012.

\bibitem{joye14}
C.~D. Joye, A.~M. Cook, J.~P. Calame, D.~K. Abe, A.~N. Vlasov, I.~A.
  Chernyavskiy, K.~T. Nguyen, E.~L. Wright, D.~E. Pershing, T.~Kimura,
  M.~Hyttinen, and B.~Levush, ``Demonstration of a high power, wideband
  220-{GHz} traveling wave amplifier fabricated by {UV-LIGA},'' \emph{IEEE
  Trans. Electron Devices}, vol.~61, no.~6, pp. 1672--1678, 2014.

\bibitem{Armstrong2018}
C.~M. Armstrong, R.~Kowalczyk, A.~Zubyk, K.~Berg, C.~Meadows, D.~Chan,
  T.~Schoemehl, R.~Duggal, N.~Hinch, R.~B. True, R.~Tobin, M.~Sweeney, and
  B.~Weatherford, ``A compact extremely high frequency {MPM} power amplifier,''
  \emph{IEEE Trans. Electron Devices}, vol.~65, no.~6, pp. 2183--2188, 2018.

\bibitem{Armstrong2020}
C.~M. Armstrong, ``These vacuum devices stood guard during the cold war,
  advanced particle physics, treated cancer patients, and made the beatles
  sound better,'' \emph{IEEE Spectrum}, vol.~57, no.~11, pp. 30--36, 2020.

\bibitem{chong13}
C.~K. Chong, D.~A. Cordrey, R.~C. Dawson, J.~W. Forster, D.~A. Layman, M.~L.
  Ramay, R.~J. Stolz, and C.~D. Washington, ``High power millimeter wave helix
  {TWT} programs at {L-3 ETI},'' in \emph{2013 IEEE 14th International Vacuum
  Electronics Conference (IVEC)}, 2013, pp. 1--2.

\bibitem{Nguyen2014}
K.~T. Nguyen, A.~N. Vlasov, L.~Ludeking, C.~D. Joye, A.~M. Cook, J.~P. Calame,
  J.~A. Pasour, D.~E. Pershing, E.~L. Wright, S.~J. Cooke, B.~Levush, D.~K.
  Abe, D.~P. Chernin, and I.~A. Chernyavskiy, ``Design methodology and
  experimental verification of serpentine/folded-waveguide twts,'' \emph{IEEE
  Trans. Electron Devices}, vol.~61, no.~6, pp. 1679--1686, 2014.

\bibitem{baig17}
A.~Baig, D.~Gamzina, T.~Kimura, J.~Atkinson, C.~Domier, B.~Popovic, L.~Himes,
  R.~Barchfeld, M.~Field, and N.~C. Luhmann, ``Performance of a {Nano-CNC}
  machine {220-GHz} traveling wave tube amplifier,'' \emph{IEEE Trans. Electron
  Devices}, vol.~64, no.~5, pp. 2390--2397, 2017.

\bibitem{mineo10}
M.~Mineo and C.~Paoloni, ``Double-corrugated rectangular waveguide slow-wave
  structure for terahertz vacuum devices,'' \emph{IEEE Trans. Electron
  Devices}, vol.~57, no.~11, pp. 3169--3175, 2010.

\bibitem{paoloni13}
C.~Paoloni, A.~Di~Carlo, F.~Bouamrane, T.~Bouvet, A.~J. Durand, M.~Kotiranta,
  V.~Krozer, S.~Megtert, M.~Mineo, and V.~Zhurbenko, ``Design and realization
  aspects of {1-THz} cascade backward wave amplifier based on double corrugated
  waveguide,'' \emph{IEEE Trans. Electron Devices}, vol.~60, no.~3, pp.
  1236--1243, 2013.

\bibitem{paoloni14}
C.~Paoloni and M.~Mineo, ``Double corrugated waveguide for {G}-band traveling
  wave tubes,'' \emph{IEEE Trans. Electron Devices}, vol.~61, no.~12, pp.
  4259--4263, 2014.

\bibitem{Basu22}
R.~Basu, J.~Gates, P.~Narasimhan, R.~Letizia, and C.~Paoloni, ``E-band
  traveling wave tube for high data rate wireless links,'' in \emph{2022 47th
  International Conference on Infrared, Millimeter and Terahertz Waves
  (IRMMW-THz)}, 2022, pp. 1--2.

\bibitem{crepeau64}
P.~Crepeau and P.~McIsaac, ``Consequences of symmetry in periodic structures,''
  \emph{Proc. IEEE}, vol.~52, no.~1, pp. 33--43, 1964.

\bibitem{Mittra65}
R.~Mittra and S.~R. Laxpati, ``Propagation in a waveguide with glide reflection
  symmetry,'' \emph{Can. J. Phys}, vol.~43, pp. 353--372, 1965.

\bibitem{Hessel73}
A.~Hessel, M.~H. Chen, R.~Li, and A.~Oliner, ``Propagation in periodically
  loaded waveguides with higher symmetries,'' \emph{Proc. IEEE}, vol.~61,
  no.~2, pp. 183--195, 1973.

\bibitem{quevedo21}
O.~Quevedo-Teruel, Q.~Chen, F.~Mesa, N.~J.~G. Fonseca, and G.~Valerio, ``On the
  benefits of glide symmetries for microwave devices,'' \emph{IEEE Journal of
  Microwaves}, vol.~1, no.~1, pp. 457--469, 2021.

\bibitem{Gould58}
R.~W. Gould, ``Characteristics of traveling-wave tubes with periodic
  circuits,'' \emph{IRE Trans. Electron Devices}, vol.~5, no.~3, pp. 186--195,
  1958.

\bibitem{Harvey60}
A.~Harvey, ``Periodic and guiding structures at microwave frequencies,''
  \emph{IRE Trans. Microwave Theory Tech.}, vol.~8, no.~1, pp. 30--61, 1960.

\bibitem{Kieburtz70}
R.~Kieburtz and J.~Impagliazzo, ``Multimode propagation on radiating
  traveling-wave structures with glide-symmetric excitation,'' \emph{IEEE
  Trans. Antennas Propag.}, vol.~18, no.~1, pp. 3--7, 1970.

\bibitem{Staprans73}
A.~Staprans, E.~McCune, and J.~A. Ruetz, ``High-power linear-beam tubes,''
  \emph{Proc. IEEE}, vol.~61, no.~3, pp. 299--330, 1973.

\bibitem{quevedo20}
O.~Quevedo-Teruel, G.~Valerio, Z.~Sipus, and E.~Rajo-Iglesias, ``Periodic
  structures with higher symmetries: Their applications in electromagnetic
  devices,'' \emph{IEEE Microw. Mag}, vol.~21, no.~11, pp. 36--49, 2020.

\bibitem{ebrahimpouri18}
M.~Ebrahimpouri, E.~Rajo-Iglesias, Z.~Sipus, and O.~Quevedo-Teruel,
  ``Cost-effective gap waveguide technology based on glide-symmetric holey
  {EBG} structures,'' \emph{IEEE Trans. Microwave Theory Tech.}, vol.~66,
  no.~2, pp. 927--934, 2018.

\bibitem{Field18}
M.~Field, T.~Kimura, J.~Atkinson, D.~Gamzina, N.~C. Luhmann, B.~Stockwell,
  T.~J. Grant, Z.~Griffith, R.~Borwick, C.~Hillman, B.~Brar, T.~Reed,
  M.~Rodwell, Y.-M. Shin, L.~R. Barnett, A.~Baig, B.~Popovic, C.~Domier,
  R.~Barchfield, J.~Zhao, J.~A. Higgins, and Y.~Goren, ``Development of a
  100-{W} 200-{GHz} high bandwidth mm-wave amplifier,'' \emph{IEEE Trans.
  Electron Devices}, vol.~65, no.~6, pp. 2122--2128, 2018.

\bibitem{Basu2021}
R.~Basu, J.~M. Rao, R.~Letizia, and C.~Paoloni, ``Offset double corrugated
  waveguide,'' in \emph{2021 22nd International Vacuum Electronics Conference
  (IVEC)}, 2021, pp. 1--2.

\bibitem{Patent}
\BIBentryALTinterwordspacing
C.~Paoloni, ``Waveguide,'' patentus US 201\,602\,845O2A1, 9 29, 2016. [Online].
  Available: \url{https://patents.google.com/patent/US20160284502A1/en}
\BIBentrySTDinterwordspacing

\bibitem{paoloni15}
C.~Paoloni, M.~Mineo, M.~Henry, and P.~G. Huggard, ``Double corrugated
  waveguide for {Ka}-band traveling wave tube,'' \emph{IEEE Trans. Electron
  Devices}, vol.~62, no.~11, pp. 3851--3856, 2015.

\bibitem{Figotin07SlowWaveRes}
A.~Figotin and I.~Vitebskiy, ``Slow-wave resonance in periodic stacks of
  anisotropic layers,'' \emph{Phys. Rev. A}, vol.~76, p. 053839, Nov 2007.

\bibitem{Chabanov08StronResTra}
A.~A. Chabanov, ``Strongly resonant transmission of electromagnetic radiation
  in periodic anisotropic layered media,'' \emph{Phys. Rev. A}, vol.~77, p.
  033811, Mar 2008.

\bibitem{Othman17ExpDem}
M.~A.~K. Othman, X.~Pan, G.~Atmatzakis, C.~G. Christodoulou, and F.~Capolino,
  ``Experimental demonstration of degenerate band edge in metallic periodically
  loaded circular waveguide,'' \emph{IEEE Trans. Microwave Theory Tech.},
  vol.~65, no.~11, pp. 4037--4045, 2017.

\bibitem{Zheng22SymthMeasDBE}
T.~Zheng, M.~Casaletti, A.~F. Abdelshafy, F.~Capolino, Z.~Ren, and G.~Valerio,
  ``Synthesis and measurement of {DBE} exceptional points in
  substrate-integrated waveguides,'' \emph{IEEE Trans. Microwave Theory Tech.},
  vol.~70, no.~10, pp. 4341--4353, 2022.

\bibitem{Othman16LowStarting}
M.~A.~K. Othman, M.~Veysi, A.~Figotin, and F.~Capolino, ``Low starting electron
  beam current in degenerate band edge oscillators,'' \emph{IEEE Trans. Plasma
  Sci}, vol.~44, no.~6, pp. 918--929, 2016.

\bibitem{Zuboraj17PropDBE}
A.~M. Zuboraj, B.~K. Sertel, and C.~J.~L. Volakis, ``Propagation of degenerate
  band-edge modes using dual nonidentical coupled transmission lines,''
  \emph{Phys. Rev. Appl.}, vol.~7, p. 064030, Jun 2017.

\end{thebibliography}

\end{document}